\documentclass[10pt, article, letterpaper, onecolumn]{IEEEtran}

\usepackage{times,amsmath,epsfig,verbatim}
\usepackage{subfigure,algorithm}
\usepackage{graphicx}
\usepackage{amssymb}
\usepackage{url}
\usepackage{color}
\usepackage{mathrsfs}
\usepackage{amsthm,amsfonts}
\usepackage{float}
\usepackage[square, comma, sort&compress, numbers]{natbib}
\usepackage{bm}
\usepackage{algorithmic}
\usepackage{bbm}
\usepackage{indentfirst}
\usepackage{multicol}

\newtheorem{definition}{{\bf Definition}}
\newtheorem{proposition}{{\bf Proposition}}
\newtheorem{assumption}{\bf Assumption}
\newtheorem{theorem}{{\bf Theorem}}
\newtheorem{corollary}{{\bf Corollary}}
\newtheorem{lemma}{{\bf Lemma}}
\newcommand*{\QEDA}{\hfill\ensuremath{\blacksquare}}

\begin{document}

\title{The Role of Data Cap in Optimal Two-Part Network Pricing}

\author{Xin Wang, Richard T. B. Ma, and Yinlong Xu \\
\thanks{The updated version serves an erratum to prior papers \cite{xin2005role,wang2015sdp,wang2016network}, which studied the role of data cap in optimal two-part tariffs.
In Section VI and V of \cite{xin2005role} and \cite{wang2016network}, and Section III and VI of \cite{wang2015sdp}, the theoretical analysis and simulation results of the optimal pricing do not hold in general.
In this present article, we combine the models from the original papers with new corrected results in Section \ref{sec:revenue optimal pricing} and \ref{sec:welfare optimal pricing}.}
\thanks{X. Wang and Y. Xu are with School of Computer Science and Technology, University of Science and Technology of China. R. T. B. Ma is with School of Computing, National University of Singapore. X. Wang's email address is yixinxa@mail.ustc.edu.cn.}
}

\maketitle
\begin{abstract}
Internet services are traditionally priced at flat rates; however, many Internet service providers (ISPs) have recently shifted towards two-part tariffs where a data cap is imposed to restrain data demand from heavy users. Although the two-part tariff could generally increase the revenue for ISPs and has been supported by the US FCC, the role of data cap and its optimal pricing structures are not well understood.
In this article, we study the impact of data cap on the optimal two-part pricing schemes for congestion-prone service markets. We model users' demand and preferences over pricing and congestion alternatives and derive the market share and congestion of service providers under a market equilibrium.
Based on the equilibrium model, we characterize the two-part structures of the revenue- and welfare-optimal pricing schemes.
Our results reveal that 1) the data cap provides a mechanism for ISPs to transition from the flat-rate to pay-as-you-go type of schemes, 
2) both the revenue and welfare objectives of the ISP will drive the optimal pricing towards usage-based schemes with diminishing data caps, and 3) the welfare-optimal tariff comprises lower fees than the revenue-optimal counterpart, suggesting that regulators might want to promote usage-based pricing but regulate the lump-sum and per-unit fees.
\end{abstract}

\section{Introduction}
Traditionally, Internet service providers (ISPs) use flat-rate pricing \cite{odlyzko2001internet} for network services, where users pay fixed monthly fees for unlimited data usage.
Flat-rate pricing was widely adopted because it was easy for ISPs to implement and was preferred by users for its simplicity.
However, with the rapid development and growing popularity of data intensity services, e.g., online video streaming and cloud-based applications, the Internet traffic keeps growing more than $50\%$ per annum \cite{labovitz2011internet}, which exposes some disadvantages of the flat-rate scheme. Because flat-rate does not count the users' data usage, ``bandwidth hogs'' \cite{nabipay2011flat} consume an unfair share of capacity and are subsidized by normal users, and ISPs cannot generate enough revenues to recoup their costs, especially for the mobile providers. As a consequence, most mobile LTE providers \cite{morgan2011pricing} and even broadband ISPs, e.g., Verizon \cite{segall2011verizon} and AT\(\&\)T \cite{taylor2011t}, start to introduce a data cap\footnote{\color{black}Wireless ISPs often offer multiple data plans with different caps. Recently, some of them add the so-called unlimited plans \cite{ATTunlimited,Verizonunlimited} that have an infinite cap. However, these plans usually restrict users' data usage, e.g., slow down their connection speed, once the usage amount exceeds a certain threshold.} and adopt a two-part tariff structure, a combination of the flat-rate and usage-based pricing. Under such a two-part scheme, additional charges are imposed if a user's data usage exceeds the data cap and the exceeded amount is charged based on a per-unit fee.

Although prior work \cite{edell1999providing,dai2013design,open2013policy,nabipay2011flat} has shown that data-capped schemes could help providers generate higher revenues than that under the flat-rate pricing and
the FCC chairman has recently backed usage-based pricing for broadband to penalize heavy Internet users \cite{schatz10fcc}, {\color{black}little is known about 1) the data cap's role and impact on the optimal two-part pricing structure which maximizes the providers' revenue or social welfare, and 2) potential regulations on the data cap for protecting social welfare from the monopoly providers.}


In this paper, we address these questions on a generic congestion-prone network service market, e.g., mobile, broadband or cloud services, and study the data cap under two-part pricing schemes. Unlike physical commodities, the quality of network services is intricately influenced by a negative network effect (or network externality): the more users access the service simultaneously, the worse performance it provides. We model this service congestion as a function of providers' capacity and their data load. We characterize users by their demand and values on data usage and analyze the market shares of providers under varying pricing and market structures. We derive the network congestion of providers under a market equilibrium.
Based on the equilibrium model, we analyze the effect of data cap on the provider's optimal revenue and pricing structure.
We also analyze and compare the revenue-optimal and welfare-optimal two-part schemes, and derive regulatory implications for protecting social welfare from monopoly providers.
Our main contributions and findings include the following.
\begin{itemize}
\item {\color{black}We model users' optimal data usages and preferences over providers  for any pricing scheme and congestion level (Section \ref{sec:model}).} We characterize the existence and uniqueness (Theorem \ref{theorem:equilibrium}) of a congestion equilibrium and show its monotonic dynamics (Theorem \ref{the:congestion market share impacted competition}) under varying pricing and market structures.

\item We analyze the impact of data cap on a provider's optimal pricing and revenue under two-part tariffs (Section \ref{sec:revenue optimal pricing}). We find that by decreasing data cap, the revenue-optimal pricing transforms from flat-rate to pay-as-you-go, during which the provider's revenue increases from a minimum to a maximum (Theorem \ref{theorem: revenue extremum}).
With heavier user demand, higher provider's capacity, or more intense market competition, the revenue objective will drive providers' pricing from flat-rate towards pay-as-you-go more strongly.
    
\item We characterize the welfare-optimal two-part tariff (Section \ref{sec:welfare optimal pricing}) and find that the optimal social welfare increases from a minimum to a maximum when data cap decreases (Theorem \ref{theorem: welfare extremum}) and the welfare-optimal pricing imposes lower fees than those under the revenue-optimal counterpart. These results suggest that regulators might want to encourage the use of two-part pricing with limited data caps, while regulating the lump-sum and per-unit fees.
\end{itemize}

We believe that our work provides new insights into the role of data cap in the optimal structure of the two-part tariff. Our results could help service providers design revenue-optimal pricing schemes and guide regulatory authorities to legislate desirable regulations.


\section{Related Work}\label{sec:related work}
Early studies of the two-part tariff came from economics. Oi \cite{oi1971disneyland} first studied the price discrimination via quantity discounts in the monopoly Disneyland market.  Calem \textit{et al.} \cite{calem1984multiproduct} examined and compared the revenue-optimal prices by a multi-product monopoly and a differentiated oligopoly. Littlechild \cite{littlechild1975two} studied the explicit characterizations of welfare-optimal pricing and the effect of consumption externalities. Scotchmer \cite{scotchmer1985two} explored the nature of Nash equilibrium among profit-maximizing shared facilities. However, all of these work were confined to a special case where the data/usage cap is set to be zero. Our work studies the impact and dynamic of data cap on the two-part pricing, which generalizes the special case.

As the demarcation between the lump-sum and usage-based fees, data cap plays a crucial role in the two-part tariff structures.
Prior work \cite{edell1999providing,dai2013design,open2013policy,nabipay2011flat,dai2013isp} demonstrated that data-capped schemes can increase providers' revenue compared with the traditional flat-rate pricing \cite{odlyzko2001internet}. Odlyzko \textit{et al.} \cite{odlyzko2012know} showed that ISPs could also reduce network congestion by imposing data caps. Our analysis and results also confirm these observations. Furthermore, we focus on understanding the impact of data cap on the structures of revenue-optimal and welfare-optimal tariffs.
The impact and optimal design of data cap have been empirically studied. In a qualitative study of households users under bandwidth caps, Chetty \textit{et al.} \cite{chetty2012you} studied how the uncertainties of user types and demand would impact the setting of data cap and operator's revenue, and proposed new tools to help users manage their caps.
Poularakis \textit{et al.} \cite{Poularakis2014pricing} proposed a framework to calculate the optimal data caps and empirically evaluated the gains of ISPs when they adopt data caps based on traffic datasets.
Unlike these efforts, our work adopts an analytical approach to characterize the desirable data cap and the optimal two-part pricing structures.

More generally, there have been several works that study the usage-based Internet pricing.
Hande \textit{et al.} \cite{hande2010pricing} characterized the economic loss due to ISPs' inability or unwillingness to price broadband access based on the time of use. Li \textit{et al.} \cite{li2009revenue} studied the optimal price differentiation under complete and incomplete information. Basar \textit{et al.} \cite{basar2002revenue} devised a revenue-maximizing pricing under varying user scales and network capacities. Shen \textit{et al.} \cite{shen2007optimal} investigated optimal nonlinear pricing policy design for a monopolistic service provider and showed that the introduction of nonlinear pricing provides a large profit improvement over linear pricing.
In this paper,  besides optimizing the revenue from the provider's perspective, we look into the welfare-optimal solution, through which we derive regulatory implications.
Chen \textit{et al.} \cite{chen2016impact} and Ma \cite{ma2016usage} also studied the revenue-optimal and welfare-optimal pricing schemes in network service markets. {\color{black}Chen \textit{et al.} \cite{chen2016impact} characterized the equilibrium allocations of bandwidth when monopoly or competitive wireless ISPs maximize their revenues or social welfare.} Ma \cite{ma2016usage} observed providers' optimal pricing strategies in congestion-prone service markets, e.g., cloud or broadband services. Both of them considered the purest form of usage-based pricing, which does not charge the lump-sum fee and has a zero usage cap, and thus is a special case of the two-part pricing structure studied in this paper. Besides the usage-based pricing, many other schemes, e.g., time-dependent pricing \cite{ha2012tube} and congestion-based pricing \cite{henderson2001congestion}, have also been adopted for network services.
{\color{black}Gizelis and Vergados conducted a survey \cite{gizelis2011survey} which classified and evaluated the wide variety of pricing schemes.}

From a modeling perspective, Chander \cite{chander1989optimal}, Reitman \cite{reitman1991endogenous}, and our work all consider the service market with congestion externalities. Chander \cite{chander1989optimal} studied the quality differentiation strategy of a monopoly provider and Reitman \cite{reitman1991endogenous} studied a multi-provider price competition. Both of them modeled the market as a continuum of non-atomic users, each of which is characterized by a quality-sensitivity parameter. However, this one-dimensional model only applies for flat-rate pricing. To faithfully characterize the utility of users under two-part tiered pricing, we establish a novel two-dimensional model that describes users by their data demand and valuation on data usage. {\color{black}Duan \textit{et al.} \cite{duan2015pricing} also built a two-dimensional user model, based on which they compared the two special cases of two-part pricing structure, whose data caps are zero and infinite, respectively. Our work studies the general case of two-part pricing structure, whose data cap can be any value. Besides, \cite{duan2015pricing} modeled the distribution of users' values as the uniform distribution. To capture the values of different user segments more truthfully, we derive our results based on more general distributions, including the uniform distribution.}

\section{Model}
\label{sec:model}

\subsection{Model of Users and Their Data Demand}\label{sec:user_model}
We model each user by two orthogonal characteristics: her average value of per-unit data usage \(v\) and
desirable data demand \(u\).
The user's data usage is measured by what she is billed, e.g., the number of bits transmitted or the amount of time being online.

We denote $q$ as the congestion level of an ISP. Given the network congestion $q$, we denote $\rho(u, q)$ as the user's achievable demand.

\begin{assumption}\label{ass:discount}
\(\rho(u,q)\colon \mathbb{R}_+\times\mathbb{R}_+ \mapsto \mathbb{R}_+\) is a continuous function, increasing in \(u\) and decreasing in \(q\).
It has an upper-bound \(\rho(u,0) = u\) and satisfies \(\displaystyle\lim_{q\rightarrow +\infty} \rho(u,q) = 0\).
\end{assumption}

Assumption \ref{ass:discount} states that a user's achievable data demand equals its desirable demand $u$ under no congestion and decreases monotonically when the network congestion $q$ becomes more severe.

Beside network congestion, the user's actual data usage also depends on her ISP's pricing. We consider an ISP that adopts a two-part tiered pricing structure \(\theta = (g, f, p)\), where $g, f$ and $p$ denote a data cap, a lump-sum service fee, and a per-unit usage fee, respectively.
Under this scheme, we denote $t(y,\theta)$ as the user's charge when $y$ units of data are consumed, defined as
\begin{equation}\label{eq:charge}
t(y,\theta) \triangleq f + p(y-g)^+.
\end{equation}

Intuitively, if a user's usage is below the data cap $g$, the ISP only collects the lump-sum fee \(f\); otherwise, extra charges are imposed on the usage above the data cap with the per-unit usage fee of \(p\).
This two-part structure is also a generalization of flat-rate, i.e., $g = +\infty$, and pay-as-you-go \cite{ma2016usage} pricing, i.e., $f=0$ and $g = 0$.

We denote $\pi(y,v,\theta)$ as the utility of a user with value \(v\) when she consumes $y$ units of data and the ISP's pricing strategy is \(\theta\), defined by 
\begin{equation}\label{eq:utility function}
\pi(y,v,\theta) \triangleq vy - t(y,\theta),
\end{equation}
i.e., the total traffic value minus the charge. We assume that users determine their optimal data usages that maximize their utilities. In other words, each user tries to solve the following problem:
\begin{equation}\label{eq:max utility}
\begin{split}
&\underset{y}{\text{Maximize}} \ \ \ \pi(y,v,\theta) = vy- t(y,\theta)\\
&\text{subject to} \ \ \ \, 0 \le y\le \rho(u,q),
\end{split}
\end{equation}
where a user's actual data usage is constrained by the achievable demand $\rho(u, q)$ under the network congestion $q$.
We define \(\phi = (u,v) \) as the type of the user and denote $y^*(\phi,\theta,q)$ as its optimal usage under ISP's pricing scheme $\theta$ and congestion $q$. By solving the optimization problem in (\ref{eq:max utility}), we derive the unique optimal solution for any user type $\phi$ as
\begin{equation}\label{eq:actual demand}
y^*(\phi,\theta,q) = \rho(u,q) - \big[\rho(u,q) - g\big]^+ \mathbf{1}_{\{v<p\}}.
\end{equation}
Intuitively, if a user's achievable demand \(\rho(u,q)\) is beyond the data cap \(g\) and her value \(v\) is lower than the per-unit usage fee \(p\), she will avoid consuming extra usage above \(g\) which would result in reducing her utility, and thus the optimal data demand equals the data cap, i.e., \(y^*=g\); otherwise, the optimal data demand equals her achievable demand, i.e., \(y^*=\rho(u,q)\).
Equation \ref{eq:actual demand} also shows that the optimal demand will increase if the value $v$ or desirable demand $u$ or data cap \(g\) increases, or the provider's per-unit fee $p$ or congestion $q$ decreases.

\subsection{Users' Preferences over Providers}
We consider a market that comprises of a set \(\mathcal{N}\) of providers.
We denote \(\bm{\theta}=(\theta_i:i\in \mathcal{N})\) and \(\mathbf{q}=(q_i:i\in \mathcal{N})\) as the pricing strategy and congestion vectors of the providers.
We define $y_i^*(\phi) \triangleq y^*(\phi,\theta_i,q_i)$ as the optimal demand of user type $\phi$ when it chooses provider $i$.
For any two providers \(i,j\in \mathcal{N}\), we denote \(i\succ_{\phi}j\) if users of type \(\phi\) prefer \(i\) over \(j\).

\begin{definition}\label{def:prefer} 
We denote $\pi_i(y,v,\theta)$ as the user's utility function when using provider $i$.
For any \(i,j \in \mathcal{N}\), \(i \succ_{\phi} j\) if and only if
1) \(\pi_i\big(y_i^*(\phi),v,\theta_i\big)>\pi_j\big(y_j^*(\phi),v,\theta_j\big)\) or 2) \(\pi_i\big(y_i^*(\phi),v,\theta_i\big) = \pi_j\big(y_j^*(\phi),v,\theta_j\big)\)  and $i$ is chosen over $j$ by the user based on any arbitrary tie-breaking condition.
\end{definition}

Based on Definition \ref{def:prefer}, we assume that users of type \(\phi\) would choose their favorite provider \(i\) that induces the highest utility under their optimal data usage, i.e., \(i \succ_{\phi} j, \forall j \in \mathcal{N} \backslash \{i\}\). However, a user's best provider might still induce negative utility, and therefore we allow users not to use any of the providers if they all induce negative utility as follows.

\begin{assumption}\label{ass:free}
There exists a provider, indexed by $0$, that sets $f_0=p_0=0$ and maintains a fixed  level of congestion $q_0$.
\end{assumption}

Assumption \ref{ass:free} states that there always exist a free provider that users can choose. When we set the congestion level \(q_0 = + \infty\), the free provider becomes a dummy provider which users could choose so as to guarantee zero utility.
Furthermore, when the congestion \(q_0 < + \infty\), the free provider can be used to characterize lower-end competitors in the market that have capacities to accommodate users with a certain congestion quality \(q_0\).  Because in practice, users might also find public options \cite{ma13public} such as free municipal WiFi and municipal broadband \cite{Gryta15} recently supported by the US FCC. To distinguish between the free provider and normal providers, we assume that the set $\mathcal{N}$ are all normal providers, i.e., $0\notin\cal N$.

Based on Definition \ref{def:prefer} and Assumption \ref{ass:free}, we denote \(\Phi_i\) as provider \(i\)'s market share, i.e., the set of user types that choose to use \(i\), defined by 
\begin{equation}\label{equation:market share}
\Phi_i(\bm{\theta},\mathbf{q},q_0) = \big\{\phi: i \succ_{\phi} j,\forall j \in \mathcal{N} \cup\{0\}\backslash \{i\} \big\}.
\end{equation}
Next, we show how providers' market shares \(\Phi_i(\bm{\theta},\mathbf{q},q_0)\) 
vary when the set of competing providers $\cal N$ or the pricing decisions \(\bm{\theta}\) change.

\begin{proposition}\label{proposition:pricing and competition effect}
For a set \(\mathcal{N}\) of providers, if two pricing decisions \(\bm{\hat{\theta}}\) and \(\bm{\theta}\) satisfy \(\hat{g}_i \ge g_i, \hat{f}_i \le f_i, \hat{p}_i \le p_i\) for some \(i\in \mathcal{N}\) and \(\hat{\theta}_j=\theta_j\) for all \(j\neq i\), we have \(\Phi_i(\bm{\theta},\mathbf{q},q_0)\subseteq \Phi_i(\bm{\hat{\theta}},\mathbf{q},q_0)\) and \(\Phi_j(\bm{\theta},\mathbf{q},q_0)\supseteq \Phi_j(\bm{\hat{\theta}},\mathbf{q},q_0), \ \forall \ j\neq i.\)
For two sets \(\mathcal{N}\) and \(\mathcal{N}'\) of providers, if \(\mathcal{N}\subseteq\mathcal{N}'\) and \((\theta'_i,q'_i) \!= (\theta_i,q_i)\) for \(\forall i\in \mathcal{N}\), we have $\Phi_i(\bm{\theta}',\mathbf{q}',q_0)\subseteq \Phi_i(\bm{\theta},\mathbf{q},q_0), \ \forall \ i\in \mathcal{N}$.
\end{proposition}

Proposition \ref{proposition:pricing and competition effect} states that under fixed levels of congestion $\mathbf{q}$, the market share of a provider would increase if the provider reduces its fees $f$ or $p$, or raises its data cap $g$, unilaterally. Meanwhile, the market share of any other provider will decrease.
It implies that monopolistic providers could use fees and data cap to trade off its market share and revenue; while oligopolistic providers could compete for market shares by decreasing their fees and increasing data caps. Proposition \ref{proposition:pricing and competition effect} also implies that bringing new providers into the market will intensify market competition and existing providers' market shares will decrease because some of their users might switch to the new providers.

Proposition \ref{proposition:pricing and competition effect} holds under the condition of fixed network congestion. However, a provider's congestion level depends on its market share and the data usage of its users. We discuss the dynamics of network congestion and equilibrium in the next subsection.

\subsection{Network Congestion and Equilibrium}\label{sec:market_equilibrium}
We denote \(U\) and \(V\) as the maximum desirable data demand and maximum per-unit data value of all users. Thus, the domain of users can be defined by $\Phi \triangleq [0,U]\times[0,V]$.
We model the distribution of users by the measure space \((\Phi,\mu)\), where \(\mu\) is a product measure
\begin{equation*}
\mu(E_1\times E_2) = \mu_u(E_1) \times \mu_v(E_2), \ \forall \ E_1 \subseteq [0,U], E_2 \subseteq [0,V],
\end{equation*}
where \(\mu_u\) and \(\mu_v\) are two continuous measures, defined by
\begin{equation*}
\begin{split}
&\mu_u\big((u_1,u_2]\big) = F_u (u_2) - F_u (u_1), \forall u_1 \le u_2, \,\text{and}\\
&\mu_v\big((v_1,v_2]\big) = F_v (v_2) - F_v (v_1), \forall v_1 \le v_2, \\
\end{split}
\end{equation*}
for non-decreasing distribution functions \(F_u(u)\) and \(F_v(v)\). The distribution functions \(F_u(u)\) and \(F_v(v)\) capture the population of users whose desirable demand and value are less than or equal to \(u\) and \(v\), respectively. Thus, \(\mu(E_1\times E_2)\) measures the population of users whose types belong to the set \(E_1\times E_2\).

Based on the distribution of users, we define the provider \(i\)'s data load, i.e., the aggregate data demand of users of \(i\), by
\begin{equation}\label{eq:data load}
\begin{split}
d_i = D(\Phi_i(\bm{\theta},\mathbf{q},q_0);\theta_i,q_i) \triangleq \int_{\Phi_i(\bm{\theta},\mathbf{q},q_0)} y_i^*(\phi,\theta_i,q_i) d\mu\\
\end{split}
\end{equation}

On the one hand, given congestion levels \(\mathbf{q}\), provider \(i\) has an induced data load \(d_i = D(\Phi_i;\theta_i,q_i)\). On the other hand, the provider's congestion level \(q_i\) is influenced by its data load \(d_i\). We denote $c_i$ as provider $i$'s capacity and model its congestion as a function $q_i = Q_i(d_i, c_i)$ of data load $d_i$ and capacity $c_i$.

\begin{assumption}\label{ass:congestion_function}
$Q_i(d_i, c_i):\mathbb{R}^2_+\mapsto \mathbb{R}_+$ is continuous, increasing in $d_i$, decreasing in $c_i$ and satisfies $Q_i(0, c_i)=0$.
\end{assumption}

Different forms of the congestion function $Q_i$ can be used to model the different technologies used by the provider. Assumption \ref{ass:congestion_function} implies that a provider \(i\)'s congestion increases (decreases) when its data load $d_i$ (capacity $c_i$) increases, and no congestion exists when no user consumes data from the provider.

We denote $Q_i^{-1}(q_i, c_i)$ as the inverse function of $Q_i(d_i,c_i)$ with respect to $d_i$, which defines the implied load under the capacity $c_i$ and an observed congestion level of $q_i$.
By Assumption \ref{ass:congestion_function}, we know that $Q_i^{-1}(q_i, c_i)$ is continuous, increasing in both $q_i$ and $c_i$, and satisfies $Q_i^{-1}(0, c_i)=0$.
We denote \(\mathbf{c}=(c_i:i\in \mathcal{N})\) as the vector of the capacities of all the providers.
Under an exogenous level \(q_0\) of congestion and when the providers make exogenous pricing decisions $\bm{\theta}$ and capacity planning decisions $\mathbf{c}$, the resulting congestion $\mathbf{q}$ of the providers can be determined endogenously when users choose their best providers.
We define such a market equilibrium of the system as follows.

\begin{definition}\label{definition:equilibrium}
Given the congestion level \(q_0\), for a set $\mathcal{N}$ of providers with any fixed pricing strategies $\bm{\theta}$ and capacities $\mathbf{c}$, $\mathbf{q}$ is an equilibrium if and only if
\begin{equation*}
q_i = Q_i\Big(D\big(\Phi_i(\bm{\theta},\mathbf{q},q_0);\theta_i,q_i\big), c_i\Big),\quad \forall i\in \mathcal{N}.
\end{equation*}
\end{definition}

To better understand the above definition, we can equivalently rephrase the above equality condition as
$D\big(\Phi_i(\bm{\theta},\mathbf{q},q_0);\theta_i,q_i\big) = Q_i^{-1}(q_i, c_i)$,
where the left-hand side is the induced data load of provider \(i\) given its market share $\Phi_i(\bm{\theta},\mathbf{q},q_0)$, pricing strategy $\theta_i$ and congestion $q_i$ and the right-hand side is its implied data load under capacity $c_i$. In equilibrium, both equal the actual aggregate user demand $d_i$. Because the equilibrium depends on the pricing strategies \(\bm{\theta}\), capacities \(\mathbf{c}\) and congestion \(q_0\), we also denote it as \(\mathbf{q}(\bm{\theta},\mathbf{c},q_0)\).

\begin{theorem}\label{theorem:equilibrium} 
Under Assumption \ref{ass:discount}-\ref{ass:congestion_function}, for any fixed pricing strategies $\bm{\theta}$, capacities $\mathbf{c}$ and congestion $q_0$, there exists at least one market equilibrium, and the equilibrium(s) must have the following properties: 
\begin{enumerate}
\item {\color{black}for the market including only two normal providers, i.e., \(\mathcal{N}=\{i,j\}\), if there exist multiple equilibria, then \((q_i-\hat{q}_i)(q_j-\hat{q}_j)\ge 0\) for any two equilibria \(\mathbf{q}\) and \(\mathbf{\hat{q}}\).}
\item for the market including only one normal provider, i.e., \(|\mathcal{N}|=1\), the equilibrium is unique.
\end{enumerate}
\end{theorem}

Theorem \ref{theorem:equilibrium} states that under minor assumptions of the demand (Assumption \ref{ass:discount}) and congestion (Assumption \ref{ass:congestion_function}), the existence of market equilibrium can be guaranteed. 
{\color{black}The first property shows that when the market has only two normal providers, the providers' congestion levels have the same changing direction (i.e., both increase or both decrease) as the market shifts from an equilibrium to another equilibrium.}
The second property tells that when the market has only one normal provider, it has a unique equilibrium. For this case, we denote \(I\) as the provider and define \(\mathcal{I} = \{I\}\). The next theorem shows how its congestion level varies when its pricing strategy $\theta_I$ and capacity $c_I$ change, or other providers enter the market.

\begin{theorem}\label{the:congestion market share impacted competition}
In a market $\cal I$, provider $I$'s unique level of congestion \(q_I(\theta_I,c_I,q_0)\) in equilibrium is non-increasing in its fees \(f_I\), \(p_I\) and capacity \(c_I\), but is non-decreasing in its data cap \(g_I\) and congestion \(q_0\) of the free provider.
When new providers enter and form a new market \(\mathcal{N} \supset \mathcal{I} \) and $I$ keeps its pricing \(\theta_I\) and capacity \(c_I\), $I$'s congestion \(q_I(\bm{\theta},\mathbf{c},q_0)\) under any new equilibrium satisfies \(q_I(\bm{\theta},\mathbf{c},q_0) \le q_I(\theta_I,c_I,q_0)\).
\end{theorem}

\begin{figure}[t]
 \centering
 \subfigure[equilibrium market share]{
 \label{fig:market share 1}
 \includegraphics[width=0.23\textwidth]{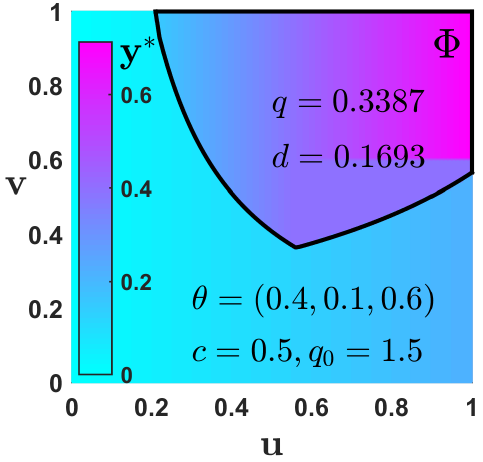}}
 \subfigure[\(c_I\) decreases to \(0.3\)]{
 \label{fig:market share 2}
 \includegraphics[width=0.23\textwidth]{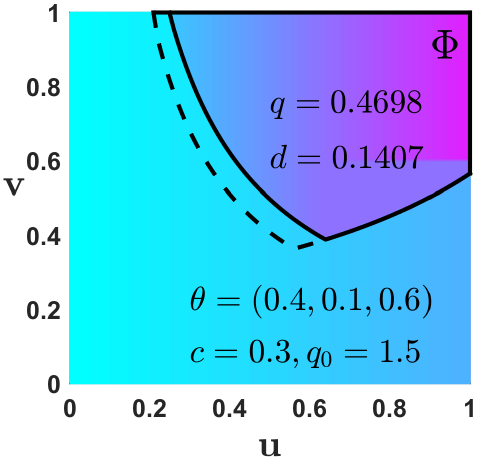}}
 \subfigure[\(q_0\) increases to \(3.0\)]{
 \label{fig:market share 3}
 \includegraphics[width=0.23\textwidth]{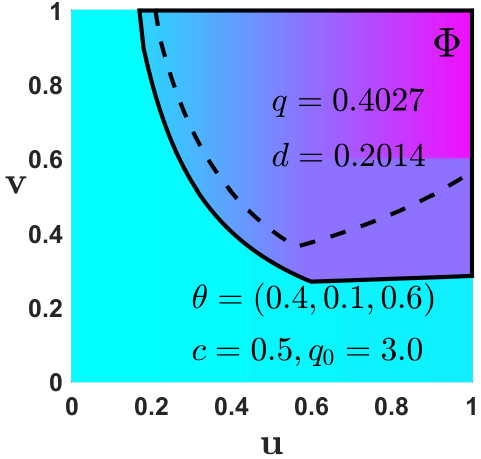}}\\
  \subfigure[\(f_I\) reduces to \(0.05\)]{
 \label{fig:market share 4}
 \includegraphics[width=0.23\textwidth]{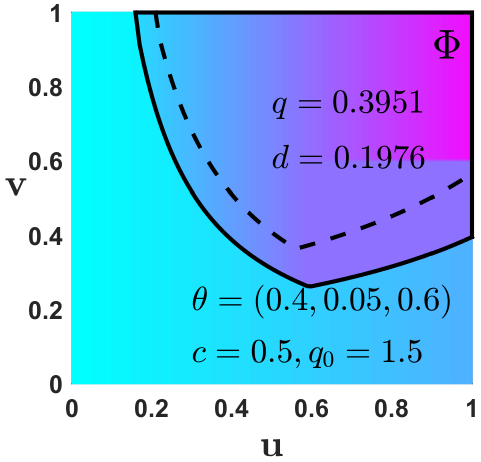}}
  \subfigure[\(p_I\) reduces to \(0.1\)]{
 \label{fig:market share 5}
 \includegraphics[width=0.23\textwidth]{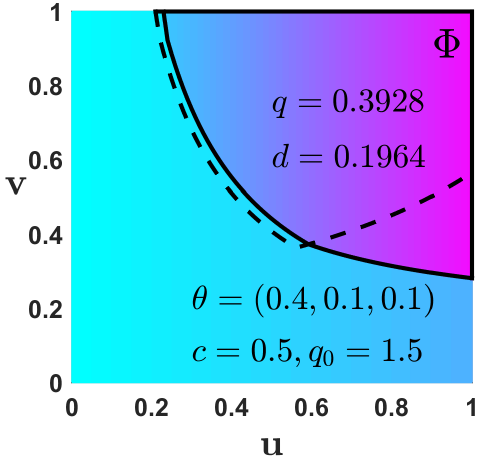}}
  \subfigure[\(g_I\) relaxes to \(0.6\)]{
 \label{fig:market share 6}
 \includegraphics[width=0.23\textwidth]{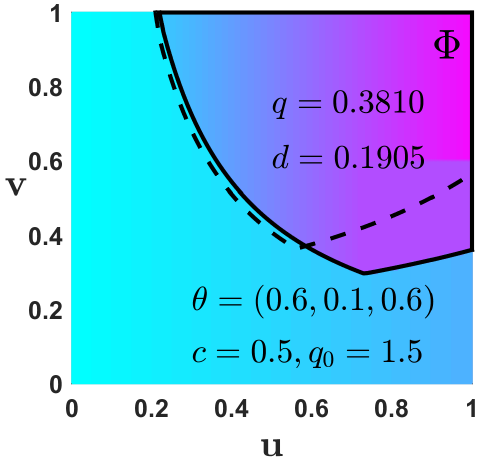}}
 \caption{Shift of market share for the provider}
 \label{fig:market share}
\end{figure}

Theorem \ref{the:congestion market share impacted competition} states that when the normal provider increases fees or decreases data cap, its induced congestion level will be alleviated because its market share and data load will decrease. When the provider increases capacity, the congestion level will also decrease, although the resulting market share and data load will increase. Furthermore, if the congestion \(q_0\) deteriorates, the provider's congestion will increase, as some users of the free provider will shift to it. It also shows that with more competing providers, the existing monopoly's congestion will decrease, as users have more choices and may switch to other providers. This implies that competition could alleviate the congestion of the providers because effectively more providers bring higher capacity to the entire market.

Under any market equilibrium $\mathbf{q}(\bm{\theta},\mathbf{c},q_0)$, we denote $R_i$ and $S_i$ as the revenue and social welfare generated by provider $i$, defined as
\begin{equation}\label{equation:welfare}
R_i \triangleq \int_{\Phi_i} t\left(y_i^*(\phi), \theta_i\right) d\mu \quad \text{and}\quad S_i \triangleq \int_{\Phi_i} v y_i^*(\phi) d\mu,
\end{equation}
where $y_i^*(\phi) = y^*(\phi,\theta_i,q_i)$ and $\Phi_i=\Phi_i(\bm{\theta},\mathbf{q},q_0)$ are evaluated at the equilibrium $\mathbf{q}(\bm{\theta},\mathbf{c},q_0)$.

\subsection{Model Parameters and Properties}
Although our model is built upon generic assumptions (Assumption \ref{ass:discount} and \ref{ass:congestion_function}), it does not yet capture the characteristics of network services.
To this end, we carefully choose the model parameters, i.e., the congestion function $Q_i(d_i, c_i)$, the achievable demand function $\rho(u,q)$ and the measure space \((\Phi,\mu)\) of user domain.

Next, we choose a quintessential form $\rho(u,q)=u  e^{-q}$ for the achievable demand function, which is used by prior work \cite{reitman1991endogenous,ma2013evolution}. Under this form, the user's achievable demand decays exponentially at a rate of $q$, i.e., the level of congestion.
We adopt the congestion function $Q_i(d_i,c_i)=d_i/c_i$, which models the {\em capacity sharing} \cite{chau2010viability} nature of network services. This form has been used in much prior work such as \cite{gibbens2000internet,chau2010viability,jain2001analysis}.

\begin{corollary}\label{corollary:equilibrium_normalization}Suppose $\rho(u,q)=u  e^{-q}$, $Q_i(d_i,c_i)=d_i/c_i$ for all \(i\in\mathcal{N}\) and let $\mathbf{q}$ be an equilibrium under parameters $q_0,\bm{\theta}, \mathbf{c}, \Phi$ and $\mu$. {\color{black}For another market with $\hat{q}_0 = q_0$, $\hat{\mathbf{f}}={\mathbf{f}}/(UV)$, $\hat{\mathbf{g}}=\mathbf{g}/U$, ${\hat{\mathbf{p}}}=\mathbf{p}/V$,  $\hat{\mathbf{c}}=\mathbf{c}/\big(U\mu(\Phi)\big)$, $\hat{\Phi} \!=\! [0,1]\times [0,1]$ and $\hat{\mu}([0,\hat{u}]\times[0,\hat{v}]) = \mu([0,u]\times[0,v])/\mu(\Phi)$ where \((\hat{u}, \hat{v}) = (u/U, v/V)\in \hat{\Phi}\), we must have $\hat{\mathbf{q}}=\mathbf{q}$ as an equilibrium under which $\hat{d_i}= d_i/\big(U\mu(\Phi)\big)$ for all $i\in \mathcal{N}$.}
\end{corollary}

Corollary \ref{corollary:equilibrium_normalization} states that under the exponential form of achievable demand and capacity sharing nature of network services, the domain and measure of users can be normalized to be $[0,1]\times[0,1]$ and probability measure, respectively. 
In particular, keeping the congestion level of the free provider unchanged, the equilibrium does not change when {\color{black}1) the lump-sum fees $\mathbf{f}$ are normalized by $UV$, 2) the data caps $\mathbf{g}$ and the users' desirable demands $u$ are normalized by $U$, 3) the per-unit fees $\mathbf{p}$ and the users' values $v$ are normalized by $V$, 4) the capacities $\mathbf{c}$ are normalized by $U\mu(\Phi)$, and 5) the user population $\mu([0,u]\times[0,v])$ is normalized by $\mu(\Phi)$.}

Based on this result, we can focus on the maximum user demand and value \(U=V=1\) (i.e., $\Phi = [0,1]\times[0,1]$) and probability distribution functions \(F_u(U)=F_v(V)=1\) (i.e., $\mu_u([0,U]) = \mu_v([0,V]) = \mu(\Phi) = 1$) without loss of generality.
Furthermore, we assume the values of users satisfies the polynomial distribution \(F_v(x) = x^{\beta}\) with a parameter \(\beta>0\), and in the later sections, we simulate the distribution of user demand by \(F_v(x) = x^{\alpha}\) with a parameter \(\alpha>0\).
In particular, when \(\alpha = 1\), user demand are uniformly distributed; otherwise, they are leaning toward the high (\(\alpha > 1\)) or low (\(\alpha < 1\)) values in the domain \([0,1]\). In summary, we will analyze the two-part tiered pricing of the providers with the following assumption.

\begin{assumption}\label{ass:model parameters} 
Any provider $i$'s congestion level satisfies $Q_i(d_i,c_i)=d_i/c_i$, users' achievable demands satisfy $\rho(u,q)=u e^{-q}$, and their values are distributed by \(F_v(x) = x^{\beta}\) for \(x\in [0,1]\).
\end{assumption}

When the level of the congestion is exogenously given, Proposition \ref{proposition:pricing and competition effect} implies that the market share of a provider will decrease when it raises fees, e.g., $f_I$ and $p_I$, or reduces data cap $g_I$.
However, the higher fees or lower data cap would alleviate the provider's congestion in equilibrium by Theorem \ref{the:congestion market share impacted competition}, which results in attracting more congestion-sensitive users to join. As a result, the dynamics of the provider's market share combines both effects and may not be monotonic as we illustrate through an example in Figure \ref{fig:market share} under Assumption \ref{ass:model parameters}.
In this example, the users are uniformly distributed, i.e., \(\alpha = \beta = 1\). In each subfigure, x-axis and y-axis vary the desirable data demand \(u\) and value \(v\) of the user types. In other words, each point in the subfigures represents a unique user type. We use different colors of points to represent the data usages \(y^*\) of user types.
In subfigure (a), the upper-right area shows the market share $\Phi_I$ of the normal provider when it makes the pricing strategy $\theta_I = (g_I, f_I, p_I) = (0.4, 0.1, 0.6)$ and capacity \(c_I=0.5\), under the congestion \(q_0=1.5\). These parameters induce congestion $q_I = 0.3387$ and data load $d_I=0.1693$ in equilibrium.
From subfigure (a) to (b), the provider's capacity $c_I$ is reduced from \(0.5\) to \(0.3\), which exacerbates the congestion level and results in a smaller market share for the provider.
From subfigure (a) to (c), the market is less competitive, i.e., $q_0$ increases from $1.5$ to $3.0$;
from subfigure (a) to (d), the provider uses a lower lump-sum fee, i.e., $f$ decreases from $0.1$ to $0.05$.
Both cases lead to a larger market share $\Phi_I$ as well as higher congestion $q_I$ and data load $d$ of the provider.
From subfigure (a) to (e) and (f), the provider decreases the per-unit fee to $p_I=0.1$ and increases the data cap to $g_I=0.6$, respectively.
The lower per-unit fee and higher data cap also induce higher data load $d_I$ and congestion $q_I$ for the provider; however, the resulting market share $\Phi_I$ attracts more low-value heavy users and loses some high-value light users.
In all cases, the congestion of the provider in equilibrium is always smaller than the congestion $q_0$ of market competitors, which is consistent with the result of Theorem \ref{theorem:equilibrium}.

\section{Revenue-Optimal Pricing}\label{sec:revenue optimal pricing}
We studied the market shares, congestion levels and data loads of providers under a market equilibrium in the previous section. In this section, we explore how providers set the two-part pricing strategy so as to optimize their revenues.
Because the data cap is the demarcation between the lump-sum (for demand below the data cap) and usage-based (for demand above the data cap) charges,
we first study the impact of data cap on the provider's optimal revenue and pricing decisions and then identify the role of data cap in the two-part pricing structure, finally we characterize the dynamics of the two-part pricing under varying system parameters, i.e., the distribution of users' demand, the capacities of the providers, the level of the market competition.

\subsection{The Impact of Data Cap}

{\color{black}We focus on a single normal provider in the market} and use the free provider with congestion \(q_0\) to model its lower-end competitors.
Under any given data cap \(g_I\), we assume the ISP chooses the lump-sum and per-unit fees, i.e., \(f_I\) and \(p_I\), to maximize its revenue \(R_I\), which is formulated by the following optimization problem.
\begin{align}\label{eq:optimization1}
&\text{Maximize} \quad R_I(g_I,f_I,p_I),\\
&\text{subject to} \quad \, f_I,p_I\ge 0.\notag
\end{align}
We denote \(f_I^*(g_I)\) and \(p_I^*(g_I)\) as an optimal solution of (\ref{eq:optimization1}), and \(R^*_I(g_I)\) as the corresponding maximum revenue. Next, we study how various data caps influence its optimal revenue $R^*_I$, and corresponding pricing decisions, i.e., $f^*_I$ and $p^*_I$.
In particular, we compare the two-part pricing with the pay-as-you-go scheme, i.e., \(f_I=g_I=0\), and the flat-rate scheme, i.e., $g_I=+\infty$. For simplification, we define the maximum revenue under the flat-rate as $R^*_{\infty} \triangleq \displaystyle\lim_{g_I\rightarrow\infty} R_I^*(g_I)$.

\begin{figure*}[t]
 \centering
 \subfigure[optimal revenue $R_I^*(g_I)$]{
 \label{fig:varying_datacap1}
 \includegraphics[width=0.256\textwidth]{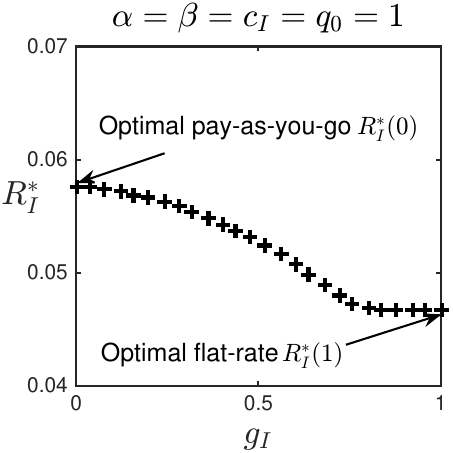}}
 \subfigure[optimal lump-sum fee $f_I^*\!(g_I\!)$]{
 \label{fig:varying_datacap2}
 \includegraphics[width=0.25\textwidth]{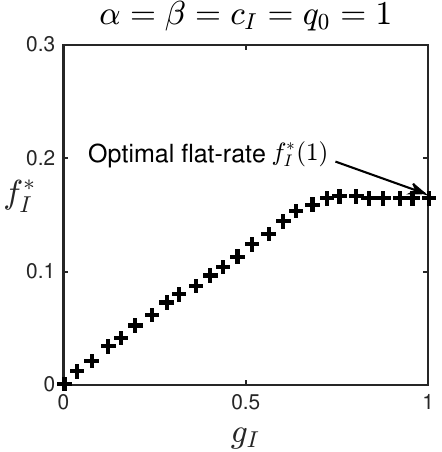}}
 \subfigure[optimal per-unit fee $p_I^*(g_I)$]{
 \label{fig:varying_datacap3}
 \includegraphics[width=0.25\textwidth]{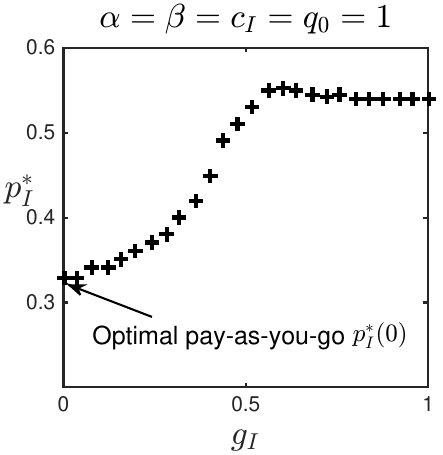}}
 \caption{Maximum revenue, optimal lump-sum and per-unit fee under varying data caps.}
 \label{fig:varying_datacap}
\end{figure*}

We consider the models under Assumption \ref{ass:model parameters} where the maximum desirable demand is normalized to $U=1$.
As a result, an effective data cap $g_I$ that might influence the users' demand has to be less than or equal to $1$.
In other words, when $g_I$ is larger than $1$, the two-part pricing is equivalent to a flat-rate scheme, i.e., $R_I^*(g_I) = R^*_{\infty} = R_I^*(1)$ for any $g_I > 1$. Thus, we will focus on $g_I\in [0,1]$ without loss of generality.
Figure \ref{fig:varying_datacap} shows how the data cap \(g_I\) influences the ISP's maximum revenue $R_I^*(g_I)$, an the optimal fees $f_I^*(g_I)$,  $p_I^*(g_I)$, under the uniform user distribution, i.e., \(\alpha=\beta=1\), and normalized ISP capacity and free provider's congestion, i.e., \(c_I=q_0=1\).
Subfigures (a) to (c) plots $R_I^*$, $f_I^*$ and  $p_I^*$ as functions of the data cap $g_I$ varying along the x-axis, respectively.

From Figure \ref{fig:varying_datacap1}, we observe that the optimal revenue \(R_I^*\) decreases monotonically with the data cap \(g_I\).
Specifically, 1) when the pricing converges to the flat-rate structure as \(g_I\) goes to one, the optimal revenue reaches a minimum; and 2) when the pricing converges to the (pure) usage-based structure as \(g_I\) tends to zero, the optimal revenue gets a maximum.
From Figures \ref{fig:varying_datacap2} and \ref{fig:varying_datacap3}, we see that the optimal fees \(f^*_I\), \(p^*_I\) both show increasing trends as the data cap \(g_I\) is relaxed.
In particular, as the data cap is zero, the lump-sum fee is also zero, implying that the optimal two-part pricing is a pay-as-you-go scheme.
The following theorem shows that our observations on the optimal revenue and fees are not particular to the model parameters, i.e., $\alpha=\beta=c_I=q_0=1$.

\begin{theorem}\label{theorem: revenue extremum} 
For any congestion \(q_0\), parameters \(\beta,c_I>0\) under Assumption 4, the provider's optimal revenue satisfies
\begin{equation*}
R^*_{\infty} \le R^*_I(g_I) \le R^*_I(0), \quad \forall \ g_I\ge 0.
\end{equation*}
Furthermore, when the data cap is zero, the revenue-optimal lump-sum fee is zero, i.e., \(f^*_I(0) = 0\).
\end{theorem}

Theorem \ref{theorem: revenue extremum} shows that the revenue generated under an optimal two-part pricing with any fixed data cap $g_I$ ($\ge 0$) is no less than that under an optimal flat-rate scheme.
In other words, an ISP could always be better off by switching from an optimal flat-rate scheme to an optimal two-part pricing scheme.
Intuitively, under any fixed lump-sum $f_I$, the two-part structure generalizes the flat-rate scheme under which $p_I = 0$ and the data cap does not play a role. Consequently, even without changing its flat-rate $f_I$, the ISP could restrict $g_I$ and increase $p_I$ to trade off between higher usage revenue from high-value users' demand and its market share, which potentially lead to higher total revenue.
This result is consistent with the views in prior work \cite{edell1999providing,dai2013design,open2013policy} that data cap could help providers extract higher revenue from the market.

Theorem \ref{theorem: revenue extremum} also states that when the data cap is zero, the revenue from the optimal two-part pricing reaches a maximum value and the corresponding lump-sum fee is zero. In other words, the optimal two-part scheme with zero data cap has a pay-as-you-go structure, which is also a global optimal two-part pricing with any data cap. Since the two-part structure is a generalization of the pay-as-you-go scheme, this result may not seem intuitive. We will provide a critical and detailed explanation of this result in the next subsection.

In summary, the data cap plays a transitional role between the flat-rate and pay-as-you-go pricing. Specifically, with the diminishing data cap, an ISP's optimal pricing transitions from a more flat-rate structure to a more pay-as-you-go structure, and the corresponding revenue changes from a minimum to a maximum.
This impact on the transformation of the optimal pricing structure holds regardless of the competition level and the ISP's capacity.

\subsection{Global Optimal Two-part Pricing Structure}\label{subsec:mechanism}

In the previous subsection, we showed that the data cap is a transitional mechanism between the flat-rate and the pay-as-you-go pricing, and the latter is a global revenue-optimal structure under the two-part pricing.
In this subsection, we provide a detailed explanation of why the pay-as-you-go pricing is a global optimal two-part pricing structure, from which we reveal that the data cap offers a way of transformation between coarse- and fine-grained pricing modes.

{\color{black} As shown in Equation (\ref{eq:actual demand}), a user's optimal data usage \(y^*\) on the ISP largely depends on her desirable demand \(u\).}
We first consider the ISP's pricing strategy on the set of the users of the same desirable demand.
Under any two-part pricing \(\theta_I = (g_I,f_I,p_I)\) and congestion level \(q_I\), 
we denote \(V_I(\theta_I,q_I,u)\) as the set of the values of the users who have desirable demand \(u\) and choose to use the ISP.
We denote \(R_I(\theta_I,q_I,u)\) and \(d_I(\theta_I,q_I,u)\) as the revenue and data load generated by these users, respectively, defined by
\begin{equation}\label{eq:demand-based revenue}
R_I(\theta_I,q_I,u) \triangleq F'_u(u) \int_{V_I(\theta_I,q_I,u)} t\big(y_I^*(u,v), \theta_I\big) dF_v
\end{equation}
\begin{equation}\label{eq:demand-based load}
d_I(\theta_I,q_I,u) \triangleq F'_u(u) \int_{V_I(\theta_I,q_I,u)} y_I^*(u,v) dF_v
\end{equation}
where $y_I^*(u,v) = y^*\big((u,v),\theta_I,q_I\big)$ and $t\big(y_I^*(u,v), \theta_I\big)$ are user's optimal data usage and charge, respectively.
To distinguish the pay-as-you-go pricing from the two-part pricing, we denote it as \(\tilde{\theta}_I \triangleq (0,0,\tilde{p}_I)\).

\begin{proposition}\label{proposition:grained comparison} 
Under Assumption \ref{ass:model parameters} and any fixed congestion levels \(q_I, q_0\) and user demand \(u\), for any two-part pricing \(\theta_I\), there always exists a pay-as-you-go pricing \(\tilde{\theta}_I\) satisfying that \(R_I(\tilde{\theta}_I,q_I,u) \!=\! R_I(\theta_I,q_I,u)\) and \(d_I(\tilde{\theta}_I,q_I,u) \!\le\! d_I(\theta_I,q_I,u)\).
\end{proposition}

Proposition \ref{proposition:grained comparison} states that under a fixed congestion level, for any two-part pricing scheme, an ISP could always find a pay-as-you-go scheme which generates the same revenue but lower data load from the set of users of the same demand.
It implies that the pay-as-you-go pricing is the most efficient two-part pricing structure for the set of users of the same demand.

To further analyze the pricing schemes in the market including users of various demands, we define a \textit{demand-based} two-part pricing, denoted as \(\theta_I(u) \triangleq (g_I(u),f_I(u),p_I(u))\) (\(u\in [0,1]\)), where \(g_I(u),f_I(u)\) and \(p_I(u)\) are continuous functions of the desirable demand \(u\).
Under this pricing strategy, a user with desirable demand \(u_0 (\in [0,1])\) is charged based on the data cap \(g_I(u_0)\), the lump-sum fee \(f_I(u_0)\) and the per-unit fee \(p_I(u_0)\).
This demand-based pricing structure is a generalization of the two-part pricing \(\theta_I = (g_I, f_I, p_I)\), i.e., \(g_I(u) = g_I, f_I(u) = f_I, p_I(u) = p_I\) for any \(u\in [0,1]\).
To distinguish the two types of pricing modes, we call \(\theta_I\) as a \textit{fixed} two-part pricing.
Similarly, we refer to \(\tilde{\theta}_I(u) \triangleq (0,0,\tilde{p}_I(u))\) and \(\tilde{\theta}_I= (0,0,\tilde{p}_I)\) as \textit{demand-based} and \textit{fixed} pay-as-you-go pricing, respectively, where the former is a generalization of the latter.

Based on Definition \ref{definition:equilibrium}, we extend the definition of the congestion equilibrium from the fixed two-part pricing to the demand-based two-part pricing, as follows.

\begin{definition}\label{definition:equilibrium expansion}
 Given the congestion level \(q_0\) of free provider, for the ISP \(I\) with any fixed demand-based two-part pricing $\theta_I(u)$ and capacity $c_I$, $q_I$ is an equilibrium if and only if \(q_I = Q_I\big(D\big(\Phi_I(\theta_I(u),q_I,q_0);\theta_I(u),q_I\big), c_I\big)\)
where \(\Phi_I\) is the set of users who choose the ISP and the function \(D\) is the aggregate data usage (load) of the users.
\end{definition}

\begin{lemma}\label{lemma:equilibrium expansion} 
Under Assumption \ref{ass:discount}-\ref{ass:congestion_function}, for any given demand-based pricing strategy $\theta_I(u)$, capacity $c_I$ and congestion $q_0$, there always exists a unique equilibrium \(q_I\), satisfying \(q_I<q_0\).
\end{lemma}

Lemma \ref{lemma:equilibrium expansion} guarantees the existence and uniqueness of the congestion equilibrium under any demand-based pricing strategy.
We denote the equilibrium under the pricing strategy \(\theta_I(u)\) as \(q_I(\theta_I(u))\).
Furthermore, we denote the provider's revenue and data load from all users of various demands in equilibrium as \(R_I\big(\theta_I(u),q_I(\theta_I(u))\big)\) and \(d_I\big(\theta_I(u),q_I(\theta_I(u))\big)\), respectively, defined by
\begin{align}\label{eq:demand-based load2}
&R_I\big(\theta_I(u),q_I(\theta_I(u))\big) = \int_0^1 R_I\big(\theta_I(u_0),q_I(\theta_I(u)),u_0\big) du_0,\notag\\
&d_I\big(\theta_I(u),q_I(\theta_I(u))\big) = \int_0^1 d_I\big(\theta_I(u_0),q_I(\theta_I(u)),u_0\big) du_0
\end{align}
where \(R_I\!\big(\theta_I(u_0),q_I(\theta_I(u)),\!u_0\big)\) and \(d_I\!\big(\theta_I(u_0),q_I(\theta_I(u)),\!u_0\big)\) are the revenue and the data load generated from the users of the desirable demand \(u_0\).

\begin{proposition}\label{proposition:grained comparison extension} 
Under Assumption \ref{ass:model parameters}, for any demand-based two-part pricing \(\theta_I(u)\), there always exists a demand-based pay-as-you-go pricing \(\tilde{\theta}_I(u)\) satisfying that \(R_I\big(\tilde{\theta}_I(u),q_I(\tilde{\theta}_I(u))\big) \ge R_I\big(\theta_I(u),q_I(\theta_I(u))\big)\).
\end{proposition}

Proposition \ref{proposition:grained comparison extension} tells that under equilibrium, for any demand-based two-part pricing scheme, there must exist a demand based pay-as-you-go pricing which could generate a higher or the same revenue from all users of various demands.
The reason is that from Proposition \ref{proposition:grained comparison}, for the given demand-based two-part pricing \(\theta_I(u)\), we could always find a demand-based pay-as-you-go pricing \(\tilde{\theta}_I(u)\) which generates the same revenue but no larger data load from users of various demands, under the same congestion level \(q_I(\theta_I(u))\).
Because the congestion level increases with the data load based on Assumption \ref{ass:congestion_function}, the congestion equilibrium \(q_I(\tilde{\theta}_I(u))\) must be no larger than the congestion \(q_I(\theta_I(u))\).
Furthermore, because the provider's revenue is non-increasing in its congestion level, the ISP's revenue under the pay-as-you-go pricing \(\tilde{\theta}_I(u)\) is no less than that under the two-part pricing \(\theta_I(u)\) in equilibrium.

Proposition \ref{proposition:grained comparison extension} implies that the demand-based pay-as-you-go pricing is an optimal demand-based two-part pricing structure. However, it is not enough to declare that the fixed pay-as-you-go and fixed two-part pricing also satisfy this relation. To achieve this, we provide the following proposition.

\begin{proposition}\label{proposition:grained explanation} 
Under Assumption \ref{ass:model parameters}, \(\tilde{\theta}_I \triangleq \big(0,0,\tilde{p}_I\big)\) is a revenue-optimal fixed pay-as-you-go pricing, if and only if \(\tilde{\theta}_I(u) \triangleq \big(0,0,\tilde{p}_I(u)\big)\) where \(\tilde{p}_I(u) = \tilde{p}_I\) for all \(u\in [0,1]\) is a revenue-optimal demand-based pay-as-you-go pricing.
\end{proposition}

Proposition \ref{proposition:grained explanation} shows that an optimal fixed pay-as-you-go pricing scheme is also an optimal demand-based pay-as-you-go pricing scheme.
Intuitively, the pay-as-you-go structure is a pure usage-based pricing mode, i.e., the charge for a user is proportional to her usage. Thus, the revenue-optimal price only depends on the distribution of user value. Because we assume that the user's value and demand are independent, under the demand-based pay-as-you-go pricing structure, the revenue-optimal prices for the user sets of different desirable demands are all the same, which is also a revenue-optimal price under the fixed pay-as-you-go pricing structure.

So far, we can provide a full explanation of why the fixed pay-as-you-go pricing is a revenue-optimal fixed two-part pricing structure. Based on Proposition \ref{proposition:grained explanation}, an optimal fixed pay-as-you-go scheme is also an optimal demand-based pay-as-you-go scheme, which generates no lower revenue compared with any demand-based two-part scheme from Proposition \ref{proposition:grained comparison extension}. Because the fixed two-part pricing is a specialization of the demand-based pricing, the optimal fixed pay-as-you-go scheme obtains no lower revenue than any fixed two-part scheme, i.e., the pay-as-you-go pricing is a revenue-optimal two-part pricing structure.

From Proposition \ref{proposition:grained comparison extension} and \ref{proposition:grained explanation}, we can also see the advantage of the pay-as-you-go pricing relative to the two-part pricing, i.e., the pay-as-you-go structure is a more fine-grained pricing mode compared to the two-part structure with a positive data cap. Specifically, the pay-as-you-go pricing is completely usage-based and fine-grained, which can achieve the revenue optimization in the market including users of various demands. By contract, the coarse-grained flat-rate has different optimal fees for users of different demands and thus it can not optimize the provider's revenue from users of various demands, which also occurs in the two-part pricing with any positive data cap, the combination of the flat-rate and the pay-as-you-go pricing.

In summary, decreasing the data cap transforms the optimal two-part pricing from a more flat-rate structure to a more pay-as-you-go structure. This shift from a more coarse-grained pricing scheme to a more fine-grained pricing scheme help the providers extracting higher revenue in the market including users of various demands.

\subsection{Dynamics of Two-part Pricing}

In the previous two subsections, we showed that the ISP's optimal revenue increases from a minimum to a maximum when data cap decreases. In this subsection, we further explore the change of the increase rate of the revenue under varying system parameters, e.g., the distribution of the user demand \(\alpha\), the capacity of the ISP \(c_I\), and the level of the market competition \(q_0\).

As shown in Figure \ref{fig:varying_datacap}, the provider's revenue increases as the data cap decreases. Furthermore, we denote \(k_I^*(g_I)\) as the increase rate of the optimal revenue under the two-part pricing with data cap \(g_I\) compared to the optimal revenue under the flat-rate, defined by
\(k_I^*(g_I) = \left[R_I^*(g_I) - R_{\infty}^*\right]/R_{\infty}^*\).
The increase rate \(k_I^*(g_I)\) reflects the growth level of the optimal revenue when the pricing changes from the flat-rate to the two-part scheme with the data cap \(g_I\).

\begin{figure}[t]
 \centering
 \subfigure[optimal revenue $R_I^*(g_I)$]{
 \label{fig:revenue_demand1}
 \includegraphics[width=0.255\textwidth]{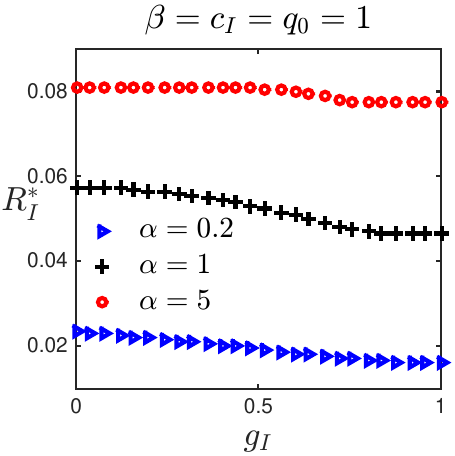}}
 \subfigure[increase rate $k_I^*(g_I)$]{
 \label{fig:revenue_demand2}
 \includegraphics[width=0.25\textwidth]{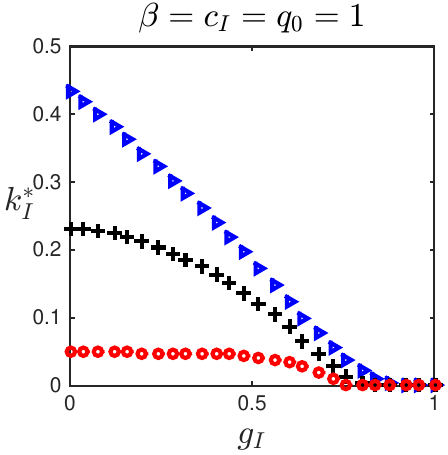}}
 \caption{Optimal revenue and increase rate under varying data caps with different values of \(\alpha\).}
 \label{fig:revenue_demand}
\end{figure}

\begin{figure}[t]
 \centering
 \subfigure[optimal revenue $R_I^*(g_I)$]{
 \label{fig:revenue_capacity1}
 \includegraphics[width=0.257\textwidth]{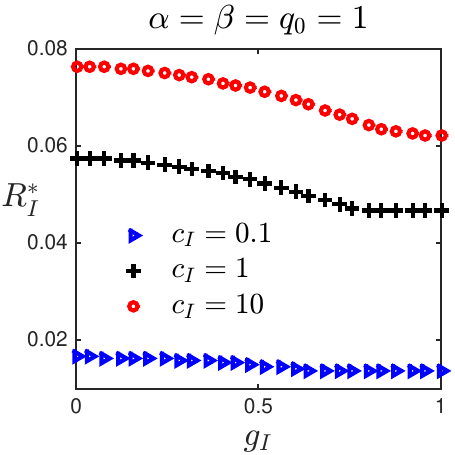}}
 \subfigure[increase rate $k_I^*(g_I)$]{
 \label{fig:revenue_capacity2}
 \includegraphics[width=0.25\textwidth]{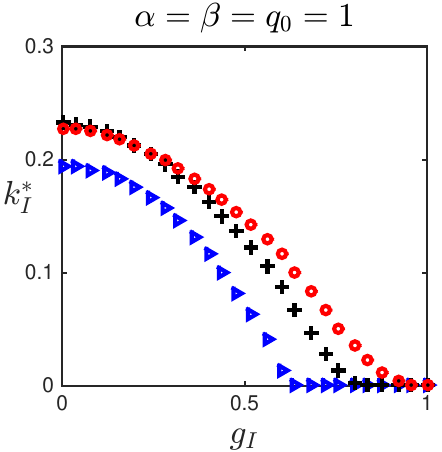}}
 \caption{Optimal revenue and increase rate under varying data caps with different values of \(c_I\).}
 \label{fig:revenue_capacity}
\end{figure}

\begin{figure}[t]
 \centering
 \subfigure[optimal revenue $R_I^*(g_I)$]{
 \label{fig:revenue_competition1}
 \includegraphics[width=0.259\textwidth]{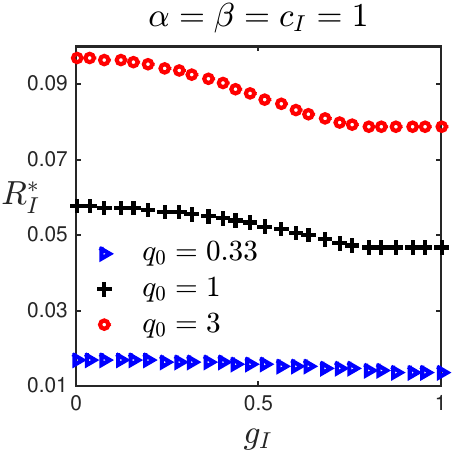}}
 \subfigure[increase rate $k_I^*(g_I)$]{
 \label{fig:revenue_competition2}
 \includegraphics[width=0.25\textwidth]{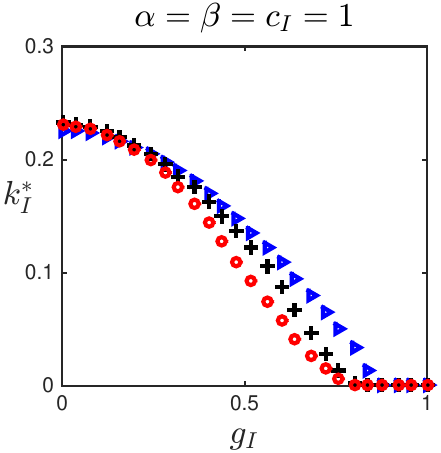}}
 \caption{Optimal revenue and increase rate under varying data caps with different values of \(q_0\).}
 \label{fig:revenue_competition}
\end{figure}

Figure \ref{fig:revenue_demand} to \ref{fig:revenue_competition} plot the ISP's optimal revenue \(R_I^*\) (left) and the increase rate \(k_I^*\) (right) as a function of its data cap \(g_I\) varying along the x-axis. Figures \ref{fig:revenue_demand}, \ref{fig:revenue_capacity} and \ref{fig:revenue_competition} show the impact of the parameter \(\alpha\) of the distribution of users' demands, the capacity \(c_I\) of the ISP, and the congestion \(q_0\) of the free provider, respectively.

Figure \ref{fig:revenue_demand} shows that small value of \(\alpha\) induces low optimal revenue \(R_I^*\) but high increase rate \(k_I^*\).
Intuitively, when \(\alpha\) is small, few users are heavy-demand and most users are light-demand. Under this case, the provider can only extract a low revenue. Because the utilities of the mostly light-demand users are small, the provider has to set a cheap lump-sum fee under the flat-rate for maximizing the revenue. However, the flat-rate structure does not restrict the users' usage, the cheap lump-sum fee can not effectively charge to the few heavy-demand users.
By contrast, when adopting the two-part pricing, the provider could optimize its revenue from the heavy-demand users by charging the per-unit fee for the usage above the data cap.
Therefore, the increase rate of the revenue under data capped schemes is high as \(\alpha\) is small.

From Figure \ref{fig:revenue_capacity}, we observe that large value of \(c_I\) leads to high optimal revenue \(R_I^*\) and increase rate \(k_I^*\).
Because when the provider's capacity \(c_I\) increases, it reduces the network congestion and increases the achievable demand of the users. Then the provider could obtain higher usage revenue from the added heavy-demand users by taking the usage-based two-part pricing, which potentially results in a faster increase rate of the total revenue.

Figure \ref{fig:revenue_competition} shows that small value of \(q_0\) induces low optimal revenue \(R_I^*\) but high increase rate \(k_I^*\).
Because when the congestion \(q_0\) of the free provider reduces, i.e., the market competition is fiercer, some low-value and light-demand users of the ISP would shift to the free provider and the ISP's optimal revenue would decrease. However, the proportion of the heavy-demand and high-value users of the ISP increase as the market competition from the free provider (lower-end competitors) becomes more intense. In this situation, the ISP can obtain higher revenue growth by adopting the usage-based two-part pricing from the heavy high-value users, and therefore the increase rate of the revenue is faster.

The above observations from Figure \ref{fig:revenue_demand} to \ref{fig:revenue_competition} also provide an explanation of why Internet service providers have shifted their pricing from the traditional flat-rate to the two-part structure recently.
In the early years of the Internet, data demands were mostly for texts and used by small groups of advanced users with high-value tasks, e.g., scientific and business purposes. The network capacities were scarce before the emergence of fiber optics backbones. Under such conditions, flat-rate pricing could be used to effectively recoup costs and the maximum revenue under two-part pricing would not be much higher than that under an optimal flat-rate scheme.
As mobile devices become a key means to access the Internet for end-users and multimedia content become more pervasive, together with the increase in network capacities, a number of ``bandwidth hogs'' \cite{nabipay2011flat} (i.e., heavy-demand users) emerge. As a result, providers now feel that flat-rate schemes cannot effectively influence users' demand and optimize their revenues; and therefore, start to adopt two-part pricing schemes.
Furthermore, implied by the observations, if the providers continue to extend their capacity and the competition from the free municipal WiFi and municipal broadband \cite{Gryta15} becomes fiercer, we would expect that providers will further reduce the lump-sum component and data cap to move closer to the pay-as-you-go pricing in the near future. In practice, the introduction of data cap in the two-part pricing also provides a means for providers to transition from the flat-rate to pay-as-you-go smoothly so that the changes will not be too abrupt for users and providers will not lose users due to the structural change of pricing schemes.

In summary, the structure of the optimal two-part pricing is largely influenced by the data cap, which plays a transitional role (mechanism) between the flat-rate and the pay-as-you-go type of pricing schemes. Specifically, diminishing the data cap transforms the optimal pricing from a more coarse-grained flat-rate structure to a more fine-grained pay-as-you-go structure which leads to higher revenue of providers. As heavy users emerge, providers' capacities grow or the competition from lower-end competitors becomes intense, revenue objectives will drive providers more strongly to shift from the flat-rate towards pay-as-you-go schemes.

\section{Welfare-optimal Pricing}\label{sec:welfare optimal pricing}

In the previous section, we studied the revenue-optimal pricing under which we showed the transitional role of data cap. However, the revenue-optimal solution does not maximize the social welfare, i.e., the total utility of the providers and their users. Although market competition in general improves the social welfare, {\color{black}which will be maximized in a perfectly competitive market under certain conditions \cite{chen2015bandwidth}, \cite[Chapter~10.1.3]{tremblay2012new},} a large deviation from the maximum welfare would more likely happen in a monopoly market. In the section, we focus on a monopoly market, i.e., the congestion level of the free provider is \(q_0=+\infty\). We characterize the impact of data cap under the welfare-optimal pricing structure. We also compare the welfare-optimal and revenue-optimal solutions of the monopoly provider and derive regulatory implications.

Under any given data cap \(g_I\), we denote \(f_I^\circ(g_I)\) and \(p_I^\circ(g_I)\) as the optimal lump-sum fee and per-unit fee that maximize the social welfare $S_I$ (defined in Equation \ref{equation:welfare}) and result in the optimal welfare $S_I^\circ(g_I)$.
To make a comparison, we denote $S_I^*(g_I)$ as the social welfare achieved under the provider's revenue-optimal fees $f_I^*(g_I)$ and $p_I^*(g_I)$.
Next, we study how various data caps influence its optimal welfare \(S_I^\circ\), and corresponding pricing decisions, i.e., \(f_I^\circ\) and \(p_I^\circ\).
In particular, we compare the two-part pricing with the pay-as-you-go scheme, i.e., \(f_I = g_I = 0\), and the flat-rate scheme, i.e., \(g_I = +\infty\). For simplification, we define the optimal social welfare under the flat-rate as \(S^\circ_\infty = \displaystyle\lim_{g_I\rightarrow \infty} S_I^\circ(g_I)\).

\begin{figure*}[t]
 \centering
 \subfigure[social welfares]{
 \label{fig:welfare_datacap1}
 \includegraphics[width=0.256\textwidth]{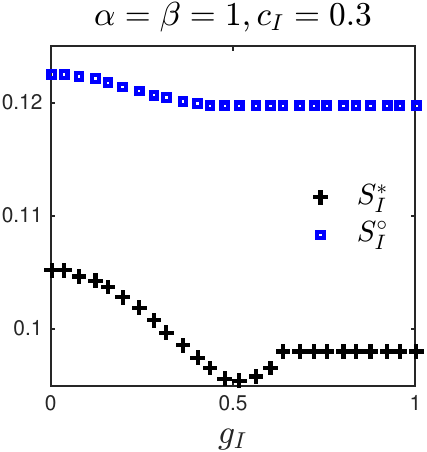}}
 \subfigure[optimal lump-sum fees]{
 \label{fig:welfare_datacap2}
 \includegraphics[width=0.25\textwidth]{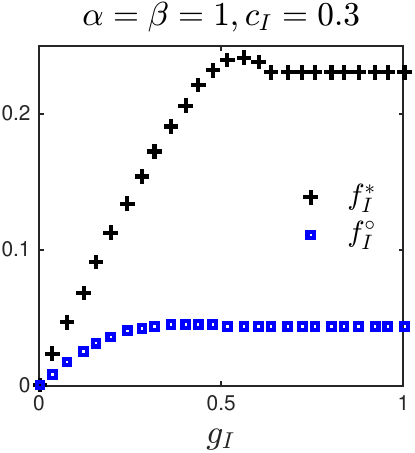}}
 \subfigure[optimal per-unit fees]{
 \label{fig:welfare_datacap3}
 \includegraphics[width=0.25\textwidth]{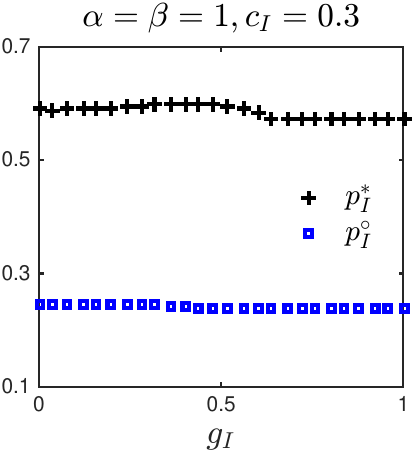}}
 \caption{Welfare-optimal and revenue-optimal pricing schemes under varying data caps.}
 \label{fig:welfare_datacap}
\end{figure*}

Figure \ref{fig:welfare_datacap} shows how data cap \(g_I\) influences the optimal social welfare \(S_I^\circ(g_I)\), and the welfare-optimal fees \(f_I^\circ, p_I^\circ\), under the uniform user distribution, i.e., \(\alpha=\beta=1\), and an ISP capacity \(c_I = 0.3\). Subfigures (a) to (c) plot \(S_I^\circ, f_I^\circ\) and \(p_I^\circ\) as functions of the data cap \(g_I\) varying along the x-axis, respectively.
As a comparison, we also plot the social welfare $S_I^*$ and the corresponding revenue-optimal fees $f_I^*, p_I^*$ in the three subfigures, respectively.
We observe that the optimal welfare \(S_I^\circ\) increases monotonically with the decrease of the data cap, which has the same changing trend as the optimal revenue \(R_I^*\) shown in Figure \ref{fig:varying_datacap}. In other words, when the pricing moves from a more flat-rate structure to a more usage-based structure with the diminishing data cap, the optimal welfare changes from a minimum to a maximum.
Further, as the data cap \(g_I\) is zero, the lump-sum fee \(f_I^\circ\) is also zero, implying the welfare-optimal pricing becomes a pay-as-you-go scheme.
The following theorem shows that these observations on the optimal welfare and fees are not particular to the model parameters, i.e., $\alpha=\beta=1$ and $c_I = 0.3$.

\begin{theorem}\label{theorem: welfare extremum} 
For any parameters \(\beta,c_I>0\) under Assumption 4, the optimal social welfare satisfies
\begin{equation*}
S^\circ_{\infty} \le S^\circ_I(g_I) \le S^\circ_I(0), \quad \forall \ g_I\ge 0.
\end{equation*}
Furthermore, when the data cap is zero, the welfare-optimal lump-sum fee is zero, i.e., \(f^\circ_I(0) = 0\).
\end{theorem}

Theorem \ref{theorem: welfare extremum} states that the social welfare generated from any optimal two-part pricing under a fixed data cap \(g_I\ge 0\) is no less than that under an optimal flat-rate scheme and no more than that under an optimal pay-as-you-go scheme.
The explanation is similar to that of the result of the optimal revenue in Theorem \ref{theorem: revenue extremum}, i.e., the pay-as-you-go structure is a pure usage-based pricing mode which is more fine-grained than the flat-rate structure. It can utilize the provider's capacity more effectively and generate higher welfare from the market including users of various demands.
Thus, with the diminishing data cap, the welfare-optimal pricing transforms from a more flat-rate structure to a more pay-as-you-go structure, and the corresponding welfare varies from a minimum to a maximum.
The results of Theorem \ref{theorem: revenue extremum} and \ref{theorem: welfare extremum} imply that both of the revenue and welfare objectives motivate the data cap decrease and move the two-part pricing towards usage-based schemes.
This implication provides justifications for regulators, e.g., the U.S. FCC, to encourage the shift to a more usage-based pricing with limited data cap for the ISPs.

From Figure \ref{fig:welfare_datacap}, we also observe that the curves of the welfare-optimal fees $f_I^{\circ}$ and $p_I^{\circ}$ are always lower than those of the revenue-optimal fees $f_I^{\ast}$ and $p_I^{\ast}$, respectively.
Because exorbitant fees reduce users' utilities and data demand, resulting in lower social welfare.
This implies that regulators might need to regulate the lump-sum and per-unit fees to guarantee higher social welfare.

In summary, our results suggest that in monopoly markets, regulators might encourage the ISPs to shift towards the two-part pricing with limited data caps; however, they need to regulate the lump-sum and per-unit fees for protecting the social welfare. {\color{black}Notice that this suggestion is complementary to the usual policy of encouraging market competition. Although competition improves the social welfare in general, there now exist many monopolistic markets where it is difficult to bring in multiple ISPs to form effective competition in a short term. For example, in many broadband markets of rural areas, ISPs have fewer incentives to deploy capacities due to low user densities and natural monopolies emerge as a result of high deployment costs. In these markets, the regulations on the data caps and prices of monopoly ISPs are strongly needed.} 

\section{Conclusions}\label{sec:conclusion}

In this paper, we study the role of data cap in the optimal structure of two-part tariffs.
We present a novel model of users' demand and preferences over pricing and congestion alternatives and derive the market share and congestion of the service providers under a market equilibrium. Based on the model, we characterize the two-part structure of the revenue-optimal and welfare-optimal pricing schemes.
We identify that the data cap provides a mechanism for ISPs to transition from the flat-rate to pay-as-you-go type of usage-based schemes.
Our results reveal that both of the revenue and welfare objectives move the two-part pricing towards usage-based schemes with the diminishing data cap.
From a perspective of regulation, our results suggest that regulators might want to promote usage-based pricing but regulate the lump-sum and per-unit fees because the welfare-optimal pricing comprises lower fees than the revenue-optimal counterpart.

\appendix
\textbf{Proof of Proposition \ref{proposition:pricing and competition effect}:}
For a set \(\mathcal{N}\) of providers and any user type \(\phi\in \Phi_i(\bm{\theta},\mathbf{q},q_0)\), by Definition \ref{def:prefer}, it satisfies that 
\begin{equation}\label{eq:proposition1_1}
\displaystyle\pi\big(y^*(\phi,\theta_i,q_i),v,\theta_i\big)\ge \pi\big(y^*(\phi,\theta_j,q_j),v,\theta_j\big), \forall j\neq i.
\end{equation}
By Equations (\ref{eq:charge}) and (\ref{eq:utility function}), the utility function \(\pi\) is non-decreasing in \(g\) and non-increasing in \(f\) and \(p\). Thus we have that 
\begin{equation}\label{eq:proposition1_2}
\pi\big(y^*(\phi,\theta_i,q_i),v,\hat{\theta}_i\big)\ge \pi\big(y^*(\phi,\theta_i,q_i),v,\theta_i\big).
\end{equation}
Because \(y^*(\phi,\hat{\theta}_i,q_i)\) is the user's optimal usage which maximizes her utility when the pricing strategy is \(\hat{\theta}_i\), it satisfies that
\begin{equation}\label{eq:proposition1_3}
\pi\big(y^*(\phi,\hat{\theta}_i,q_i),v,\hat{\theta}_i\big)\ge \pi\big(y^*(\phi,\theta_i,q_i),v,\hat{\theta}_i\big).
\end{equation}
By Inequalities (\ref{eq:proposition1_1}), (\ref{eq:proposition1_2}), and (\ref{eq:proposition1_3}), we can derive that
\begin{equation}\label{eq:proposition1_4}
\pi\big(y^*(\phi,\hat{\theta}_i,q_i),v,\hat{\theta}_i\big) \ge \pi\big(y^*(\phi,\theta_i,q_i),v,\theta_i\big) \ge \pi\big(y^*(\phi,\theta_j,q_j),v,\theta_j\big) = \pi\big(y^*(\phi,\hat{\theta}_j,q_j),v,\hat{\theta}_j\big), \forall j\neq i.
\end{equation}
By Definition \ref{def:prefer}, we know that \(\phi\in \Phi_i(\bm{\hat{\theta}},\mathbf{q},q_0)\) and therefore \(\Phi_i(\bm{\theta},\mathbf{q},q_0) \subseteq \Phi_i(\bm{\hat{\theta}},\mathbf{q},q_0)\). Similarly, we can know that \(\Phi_j(\bm{\hat{\theta}},\mathbf{q},q_0)\subseteq \Phi_j(\bm{\theta},\mathbf{q},q_0), \ \forall \ j\neq i\).

When there exist two sets \(\mathcal{N}\) and \(\mathcal{N}'\) of providers, for any provider \(i\in\mathcal{N}\subseteq \mathcal{N}'\) and any user type \(\phi\in \Phi_i(\bm{\theta}',\mathbf{q}',q_0)\), it satisfies that
\[\displaystyle\pi\big(y^*(\phi,\theta_i,q_i),v,\theta_i\big) = \pi\big(y^*(\phi,\theta'_i,q'_i),v,\theta'_i\big) \ge \pi\big(y^*(\phi,\theta'_j,q'_j),v,\theta'_j\big) = \pi\big(y^*(\phi,\theta_j,q_j),v,\theta_j\big), \forall j \in \mathcal{N} \backslash \{i\}\]
by Definition \ref{def:prefer}. Thus we have \(\phi\in \Phi_i(\bm{\theta},\mathbf{q},q_0)\) by Definition \ref{def:prefer} and therefore $\displaystyle\Phi_i(\bm{\theta}',\mathbf{q}',q_0)\subseteq \Phi_i(\bm{\theta},\mathbf{q},q_0),\forall i\in \mathcal{N}$.
\QEDA\\

\textbf{Proof of Theorem \ref{theorem:equilibrium}:} We first prove the existence of equilibrium. By Definition \ref{definition:equilibrium}, \(\mathbf{q}\) is an equilibrium if and only if \(q_i = Q_i\big(D\big(\Phi_i(\bm{\theta},\mathbf{q},q_0);\theta_i,q_i\big), c_i\big)\) for \(\forall i\in \mathcal{N}\). Since \(\bm{\theta}\), \(\mathbf{c}\) and \(q_0\) are constants, we omit them in the notation and write the above in a matrix form as \(\mathbf{q}= Q\big(D(\mathbf{q})\big)=Q\circ D(\mathbf{q})\). Thus, we can view the composite function \(Q\circ D\) as a mapping from the convex set \(\mathbb{R}_{+}^{|\mathcal{N}|}\) to itself. By Assumption \ref{ass:congestion_function}, we know that each \(Q_i(d_i,c_i)\) is continuous in \(d_i\) and thus each \(Q_i\big(D\big(\Phi_i(\bm{\theta},\mathbf{q},q_0);\theta_i,q_i\big), c_i\big)\) is continuous in \(\mathbf{q}\). To this end, we know that \(Q\circ D(\mathbf{q})\) is continuous in \(\mathbf{q}\). By Brouwer fixed-point theorem, there always exists a fixed point that satisfies \(Q\circ D(\mathbf{q})=\mathbf{q}\) and thus is an equilibrium.

Next, we prove the first property by contradiction. Suppose there exist equilibria \(\mathbf{q}\) and \(\mathbf{\hat{q}}\) that makes \((q_i-\hat{q}_i)(q_j-\hat{q}_j)< 0\). Without loss of generality, we can assume that \(q_i < \hat{q}_i\) and \(q_j > \hat{q}_j\). 
For any user type \(\phi\in \Phi_i(\bm{\theta},\hat{\mathbf{q}},q_0)\), by Definition \ref{def:prefer}, it satisfies that
\begin{equation}\label{eq:theorem1_1}
\pi\big(y^*(\phi,\theta_i,\hat{q}_i),v,\theta_i\big)\ge \pi\big(y^*(\phi,\theta_j,\hat{q}_j),v,\theta_j\big).
\end{equation}
By Assumption \ref{ass:discount}, we have \(\rho(u,q_i)>\rho(u,\hat{q}_i)\) and \(\rho(u,q_j)<\rho(u,\hat{q}_j)\). By the optimization problem in (\ref{eq:max utility}), we know that 
\begin{equation}\label{eq:theorem1_3}
\pi\big(y^*(\phi,\theta_i,q_i),v,\theta_i\big)\ge \pi\big(y^*(\phi,\theta_i,\hat{q}_i),v,\theta_i\big)\quad \text{and} \quad \pi\big(y^*(\phi,\theta_j,q_j),v,\theta_j\big)\le \pi\big(y^*(\phi,\theta_j,\hat{q}_j),v,\theta_j\big).
\end{equation}
By Inequalities (\ref{eq:theorem1_1}) and (\ref{eq:theorem1_3}), we can derive that 
\begin{equation*}\label{eq:theorem1_4}
\pi\big(y^*(\phi,\theta_i,q_i),v,\theta_i\big)\ge \pi\big(y^*(\phi,\theta_j,q_j),v,\theta_j\big).
\end{equation*}
Thus we know that \(\phi\in \Phi_i(\bm{\theta},\mathbf{q},q_0)\) by Definition \ref{def:prefer} and therefore \(\Phi_i(\bm{\theta},\hat{\mathbf{q}},q_0)\subseteq \Phi_i(\bm{\theta},\mathbf{q},q_0)\). 
By Equation (\ref{eq:actual demand}), the optimal usage \(y^*(\phi,\theta,q)\) is non-increasing in \(q\) and thus it satisfies that 
\begin{equation}\label{eq:theorem1_2}
y^*(\phi,\theta_i,q_i)\ge y^*(\phi,\theta_i,\hat{q}_i).
\end{equation}
By Inequality (\ref{eq:theorem1_2}), it satisfies that 
\[D\big(\Phi_i(\bm{\theta},\mathbf{q},q_0);\bm{\theta},\mathbf{q}\big) = \int_{\Phi_i(\bm{\theta},\mathbf{q},q_0)} y_i^*(\phi,\theta_i,q_i) d\mu \ge \int_{\Phi_i(\bm{\theta},\hat{\mathbf{q}},q_0)} y_i^*(\phi,\theta_i,\hat{q}_i) d\mu = D\big(\Phi_i(\bm{\theta},\hat{\mathbf{q}},q_0);\bm{\theta},\hat{\mathbf{q}}\big).\]
By Assumption \ref{ass:congestion_function}, we have 
\[q_i = Q\Big(D\big(\Phi_i(\bm{\theta},\mathbf{q},q_0);\bm{\theta},\mathbf{q}\big),c_i\Big) \ge Q\Big(D\big(\Phi_i(\bm{\theta},\hat{\mathbf{q}},q_0);\bm{\theta},\hat{\mathbf{q}}\big),c_i\Big) = \hat{q}_i\] 
which is contradictory with the supposition of \(q_i < \hat{q}_i\).

Finally, we prove the second property by contradiction. Suppose there exist two equilibria \(q_I\) and \(\hat{q}_I\) under fixed pricing strategy \(\theta_I\), capacity \(c_I\) and congestion \(q_0\). Without loss of generality, we assume that \(q_I<\hat{q}_I\). 
For any user type \(\phi\in \Phi_I(\theta_I,\hat{q}_I,q_0)\), by Definition \ref{def:prefer}, it satisfies that
\begin{equation}\label{eq:theorem1_5} 
\pi\big(y^*(\phi,\theta_I,\hat{q}_I),v,\theta_I\big)\ge \pi\big(y^*(\phi,\theta_0,q_0),v,\theta_0\big)
\end{equation}
where \(\theta_0=(g_0,0,0)\) is the pricing strategy of the free provider.
By Assumption \ref{ass:discount}, we have \(\rho(u,q_I)>\rho(u,\hat{q}_I)\). By the optimization problem in (\ref{eq:max utility}), we know that 
\begin{equation}\label{eq:theorem1_7}
\pi\big(y^*(\phi,\theta_I,q_I),v,\theta_I\big)\ge \pi\big(y^*(\phi,\theta_I,\hat{q}_I),v,\theta_I\big).
\end{equation} 
By Inequalities (\ref{eq:theorem1_5}) and (\ref{eq:theorem1_7}), we can derive that 
\begin{equation*}\label{eq:theorem1_8}
\pi\big(y^*(\phi,\theta_I,q_I),v,\theta_I\big)\ge \pi\big(y^*(\phi,\theta_0,q_0),v,\theta_0\big).
\end{equation*}
Thus we know that \(\phi\in \Phi_I(\theta_I,q_I,q_0)\) by Definition \ref{def:prefer} and therefore \(\Phi_I(\theta_I,\hat{q}_I,q_0)\subseteq \Phi_I(\theta_I,q_I,q_0)\). 
By Equation (\ref{eq:actual demand}), the optimal usage \(y^*(\phi,\theta,q)\) is non-increasing in \(q\) and thus it satisfies that 
\begin{equation}\label{eq:theorem1_6}
y^*(\phi,\theta_I,q_I)\ge y^*(\phi,\theta_I,\hat{q}_I).
\end{equation}
Therefore, by Inequality (\ref{eq:theorem1_6}), it satisfies that 
\[D\big(\Phi_I(\theta_I,q_I,q_0);\theta_I,q_I\big) = \int_{\Phi_I(\theta_I,q_I,q_0)} y^*(\phi,\theta_I,q_I) d\mu\ge \int_{\Phi_I(\theta_I,\hat{q}_I,q_0)} y^*(\phi,\theta_I,\hat{q}_I) d\mu = D\big(\Phi_I(\theta_I,\hat{q}_I,q_0);\theta_I,\hat{q}_I\big).\]
Furthermore, by Assumption \ref{ass:congestion_function}, we have 
\[q_I = Q\Big(D\big(\Phi_I(\theta_I,q_I,q_0);\theta_I,q_I\big),c_I\Big) \ge Q\Big(D\big(\Phi_I(\theta_I,\hat{q}_I,q_0);\theta_I,\hat{q}_I\big),c_I\Big) = \hat{q}_I\] which is contradictory with the supposition of \(q_I < \hat{q}_I\).
\QEDA\\

\textbf{Proof of Theorem \ref{the:congestion market share impacted competition}:} We first prove the congestion in equilibrium is non-increasing in the lump-sum fee by contradiction. Suppose the congestion in equilibrium is not non-increasing in the lump-sum fee, there must exist pricing strategies \(\theta_I = (g_I,f_I,p_I)\) and \(\hat{\theta}_I = (g_I,\hat{f}_I,p_I)\) satisfying \(f_I> \hat{f}_I\) and \(q_I(\theta_I,c_I,q_0)> q_I(\hat{\theta}_I,c_I,q_0)\).
By Inequality (\ref{eq:proposition1_4}), we have that for any user type \(\phi\in \Phi_I\big(\theta_I,q_I(\theta_I,c_I,q_0), q_0\big)\),
\begin{equation}\label{eq:theorem2_1}
\pi\Big(y^*\big(\phi,\hat{\theta}_I,q_I(\theta_I,c_I,q_0)\big),v,\hat{\theta}_I\Big) \ge \pi\Big(y^*\big(\phi,\theta_I,q_I(\theta_I,c_I,q_0)\big),v,\theta_I\Big).
\end{equation}
By Assumption \ref{ass:discount}, we have \(\rho\big(u,q_I(\theta_I,c_I,q_0)\big) < \rho\big(u,q_I(\hat{\theta}_I,c_I,q_0)\big)\). Thus, by the optimization problem in (\ref{eq:max utility}), we know that
\begin{equation}\label{eq:theorem2_2}
\pi\Big(y^*\big(\phi,\hat{\theta}_I,q_I(\hat{\theta}_I,c_I,q_0)\big),v,\hat{\theta}_I\Big) \ge \pi\Big(y^*\big(\phi,\hat{\theta}_I,q_I(\theta_I,c_I,q_0)\big),v,\hat{\theta}_I\Big).
\end{equation}
By Inequalities (\ref{eq:theorem2_1}) and (\ref{eq:theorem2_2}), it satisfies that
\begin{equation*}\label{eq:theorem2_3}
\pi\Big(y^*\big(\phi,\hat{\theta}_I,q_I(\hat{\theta}_I,c_I,q_0)\big),v,\hat{\theta}_I\Big) \ge \pi\Big(y^*\big(\phi,\theta_I,q_I(\theta_I,c_I,q_0)\big),v,\theta_I\Big).
\end{equation*}
Thus we know that \(\phi\in \Phi_I\big(\hat{\theta}_I,q_I(\hat{\theta}_I,c_I,q_0), q_0\big)\) by Definition \ref{def:prefer} and \(\Phi_I\big(\theta_I,q_I(\theta_I,c_I,q_0), q_0\big) \subseteq \Phi_I\big(\hat{\theta}_I,q_I(\hat{\theta}_I,c_I,q_0), q_0\big)\).
By Equation (\ref{eq:actual demand}), the optimal usage \(y^*\) is non-increasing in \(q\) and thus it satisfies that 
\begin{equation}\label{eq:theorem2_4}
y^*\big(\phi,\hat{\theta}_I,q_I(\hat{\theta}_I,c_I,q_0)\big)\ge y^*\big(\phi,\theta_I,q_I(\theta_I,c_I,q_0)\big).
\end{equation}
By Inequality (\ref{eq:theorem2_4}), we can derive that
\begin{align*}
& D\Big(\Phi_I\big(\hat{\theta}_I,q_I(\hat{\theta}_I,c_I,q_0), q_0\big);\hat{\theta}_I,q_I(\hat{\theta}_I,c_I,q_0)\Big) = \int_{\Phi_I\big(\hat{\theta}_I,q_I(\hat{\theta}_I,c_I,q_0), q_0\big)} y^*\big(\phi,\hat{\theta}_I,q_I(\hat{\theta}_I,c_I,q_0)\big) d\mu\\
& \ge \int_{\Phi_I\big(\theta_I,q_I(\theta_I,c_I,q_0), q_0\big)} y^*\big(\phi,\theta_I,q_I(\theta_I,c_I,q_0)\big) d\mu = D\Big(\Phi_I\big(\theta_I,q_I(\theta_I,c_I,q_0), q_0\big);\theta_I,q_I(\theta_I,c_I,q_0)\Big).
\end{align*}
Furthermore, by Assumption \ref{ass:congestion_function}, we have 
\begin{align*}
& q_I(\hat{\theta}_I,c_I,q_0) = Q\bigg(D\Big(\Phi_I\big(\hat{\theta}_I,q_I(\hat{\theta}_I,c_I,q_0), q_0\big);\hat{\theta}_I,q_I(\hat{\theta}_I,c_I,q_0)\Big),c_I\bigg)\\ 
& \ge Q\bigg(D\Big(\Phi_I\big(\theta_I,q_I(\theta_I,c_I,q_0), q_0\big);\theta_I,q_I(\theta_I,c_I,q_0)\Big),c_I\bigg) = q_I(\theta_I,c_I,q_0)
\end{align*}
which is contradictory with the supposition of \(q_I(\theta_I,c_I,q_0) > q_I(\hat{\theta}_I,c_I,q_0)\).
Therefore, the congestion \(q_I(\theta_I,c_I,q_0)\) in equilibrium is non-increasing in the lump-sum fee \(f_I\).

Similarly, we can prove that \(q_I(\theta_I,c_I,q_0)\) is non-increasing in the per-unit fee \(p_I\) and the capacity \(c_I\), but non-decreasing in the data cap \(g_I\) and the congestion level \(q_0\). In a similar way, we can also prove that when new providers enter the market, the existing provider's congestion level will not increase, i.e., \(q_I(\bm{\theta},\mathbf{c},q_0) \le q_I(\theta_I,c_I,q_0)\).
\QEDA

\textbf{Proof of Corollary \ref{corollary:equilibrium_normalization}:} We first prove that under \(\hat{\mathbf{q}} = \mathbf{q}\), it satisfies \(\hat{d}_i = d_i/\big(U\mu(\Phi)\big)\) for \(\forall i\in\mathcal{N}\). We denote the market share and pricing strategy of the provider \(i\) in the normalized system by \(\hat{\Phi}_i\) and \(\hat{\theta}_i = (\hat{g}_i,\hat{f}_i,\hat{p}_i)\), respectively. For any user type \(\phi = (u,v)\in \Phi\), we denote the corresponding type in the normalized system by \(\hat{\phi} = (\hat{u},\hat{v}) = (u/U,v/V) \in \hat{\Phi}\). By Equation (\ref{eq:actual demand}) and the condition of \(\rho(u,q) = u e^{-q}\), we know that
\begin{align}\label{eq:corollary1_1}
& y^*(\hat{\phi},\hat{\theta}_i,\hat{q}_i) = \hat{u}e^{-\hat{q}_i} - (\hat{u}e^{-\hat{q}_i} - \hat{g}_i)^+ \mathbf{1}_{\{\hat{v}<\hat{p}_i\}}
=\frac{ue^{-q_i}}{U} - \left(\frac{ue^{-q_i}}{U} - \frac{g_i}{U}\right)^+ \mathbf{1}_{\left\{\frac{v}{V}<\frac{p_i}{V}\right\}}\notag\\
& = \frac{1}{U}\left[ue^{-q_i} - \left(ue^{-q_i} - g_i\right)^+ \mathbf{1}_{\{v<p_i\}}\right] = \frac{y^*(\phi,\theta_i,q_i)}{U}.
\end{align}
By Equations (\ref{eq:charge}), (\ref{eq:utility function}), and (\ref{eq:corollary1_1}), it satisfeis that
\begin{align}\label{eq:corollary1_2}
&\pi\left(y^*(\hat{\phi},\hat{\theta}_i,\hat{q}_i),\hat{v},\hat{\theta}_i\right)  = \hat{v}y^*(\hat{\phi},\hat{\theta}_i,\hat{q}_i) - \hat{f}_i - \hat{p}_i\left[y^*(\hat{\phi},\hat{\theta}_i,\hat{q}_i)-\hat{g}_i\right]^+\notag\\
&= \frac{v}{V}\frac{y^*(\phi,\theta_i,q_i)}{U} - \frac{f_i}{UV} - \frac{p_i}{V}\left[\frac{y^*(\phi,\theta_i,q_i)}{U}-\frac{g_i}{U}\right]^+ =\frac{\pi\big(y^*(\phi,\theta_i,q_i),v,\theta_i\big)}{UV} 
\end{align}
implying that \(\hat{\phi}\in \hat{\Phi}_i(\hat{\bm{\theta}},\hat{\mathbf{q}},\hat{q}_0)\) if and only if \(\phi\in \Phi_i(\bm{\theta},\mathbf{q},q_0)\) by Definition \ref{def:prefer}.
Furthermore, by Equations (\ref{eq:corollary1_1}) and (\ref{eq:corollary1_2}), we can deduce that 
\begin{equation}\label{eq:corollary1_3}
\hat{d}_i = \int_{\hat{\Phi}_i(\hat{\bm{\theta}},\hat{\mathbf{q}},\hat{q}_0)} y^*(\hat{\phi},\hat{\theta}_i,\hat{q}_i) d\hat{\mu} = \int_{\Phi_i(\bm{\theta},\mathbf{q},q_0)} \frac{y^*(\phi,\theta_i,q_i)}{U} d\left(\frac{\mu}{\mu(\Phi)}\right) = \frac{d_i}{U\mu(\Phi)}.
\end{equation}

Next, we show that \(\hat{\mathbf{q}} = \mathbf{q}\) is an equilibrium in the normalized system. By Equation (\ref{eq:corollary1_3}) and the condition of \(Q_i(d,c) = d/c\), we have that
\[\hat{q}_i = q_i = Q_i(d_i,c_i) = Q_i\left(U\mu(\Phi)\hat{d}_i, U\mu(\Phi)\hat{c}_i\right) = Q_i(\hat{d}_i, \hat{c}_i)\]
implying that \(\hat{\mathbf{q}} = \mathbf{q}\) is an equilibrium in the normalized system by Definition \ref{definition:equilibrium}.
\QEDA\\

\textbf{Proof of Theorem \ref{theorem: revenue extremum}:} To understand the proof of this theorem, the notations and results in Section \ref{subsec:mechanism} are needed. 

We first prove that \(R^*_I(g_I)\ge R^*_{\infty}\) for \(\forall g_I\ge 0.\) 
We use \(\displaystyle\lim_{\hat{g}_I\rightarrow +\infty} f_I^*(\hat{g}_I)\) to denote an optimal lump-sum fee under the flat-rate, i.e., the data cap is infinite. We use \(R_I(g_I,f_I,p_I)\) to denote the ISP's revenue under the pricing strategy \(\theta_I = (g_I,f_I,p_I)\). By the optimization problem in (\ref{eq:optimization1}), we have that for \(\forall g_I\ge 0,\)
\[R_I^*(g_I) \ge R_I\left(g_I,\lim_{\hat{g}_I\rightarrow +\infty} f_I^*(\hat{g}_I),0\right) = \lim_{\hat{g}_I\rightarrow +\infty}  R_I\big(\hat{g}_I,f_I^*(\hat{g}_I),0\big) = R^*_{\infty}.\]

We then prove that \(R^*_I(g_I)\le R^*_I(0)\) for \(\forall g_I\ge 0.\) To prove this result, we only need to show that any revenue-optimal fixed pay-as-you-go pricing is also a revenue-optimal fixed two-part pricing, which in fact can be implied by Proposition \ref{proposition:grained comparison extension} and \ref{proposition:grained explanation}. In particular, by Proposition \ref{proposition:grained explanation}, a revenue-optimal fixed pay-as-you-go pricing is also a revenue-optimal demand-based pay-as-you-go pricing, which generates no lower revenue than any demand-based two-part pricing by Proposition \ref{proposition:grained comparison extension}. Because the fixed two-part pricing structure is a specialization of the demand-based two-part pricing, any revenue-optimal fixed pay-as-you-go pricing generates no lower revenue than any fixed two-part pricing. Thus, any revenue-optimal fixed pay-as-you-go pricing is also a revenue-optimal fixed two-part pricing and therefore \(R^*_I(g_I)\le R^*_I(0)\) for \(\forall g_I\ge 0.\)

We finally prove that \(f^*_I(0)=0\) by contradiction. Suppose there exists a pricing strategy \(\theta_I = (0,f_I,p_I)\) (\(f_I>0\)) satisfying that \(R_I(0,f_I,p_I) = R^*_I(0)\). We conduct a demand-based pricing strategy \(\tilde{\theta}_I(u) = \left(0,0,f_I/\rho\big(u,q_I(\theta_I)\big)+p_I\right)\). by Definition \ref{definition:equilibrium expansion}, \(q_I(\theta_I)\) is an equilibrium under the strategy \(\tilde{\theta}_I(u)\). Furthermore, it satisfies that
\begin{equation}\label{eq:theorem3_1}
R_I\left(\tilde{\theta}_I(u),q_I\big(\tilde{\theta}_I(u)\big)\right) = R_I\left(\tilde{\theta}_I(u),q_I(\theta_I)\right) = R_I(\theta_I,q_I) = R_I(0,f_I,p_I).
\end{equation}
However, because \(f_I/\rho\big(u,q_I(\theta_I)\big)+p_I\) is not a constant function of \(u\), by Proposition \ref{proposition:grained explanation}, \(\tilde{\theta}_I(u)\) must not be an revenue-optimal demand-based pay-as-you-go pricing and generates lower revenue than any revenue-optimal fixed pay-as-you-go pricing. Therefore, we have that 
\begin{equation}\label{eq:theorem3_2}
R_I\left(\tilde{\theta}_I(u),q_I\big(\tilde{\theta}_I(u)\big)\right)<R^*_I(0).
\end{equation}
by Inequalities (\ref{eq:theorem3_1}) and (\ref{eq:theorem3_2}), we know that \(R_I(0,f_I,p_I) < R^*_I(0)\), which is contradictory with the supposition of \(R_I(0,f_I,p_I) = R^*_I(0)\). Thus we know that \(f^*_I(0)\) must be zero.
\QEDA\\

\textbf{Proof of Proposition \ref{proposition:grained comparison}:} For any fixed parameters \(q_I,q_0,u\) and two-part pricing \(\theta_I\), we conduct the pay-as-you-go pricing \(\tilde{\theta}_I = (0,0,\tilde{p}_I)\) satisfying \(R_I(\tilde{\theta}_I,q_I,u) = R_I(\theta_I,q_I,u)\) and \(d_I(\tilde{\theta}_I,q_I,u) \le d_I(\theta_I,q_I,u)\) under the following two cases.

The first case is when \(q_I\ge q_0\) or \(\rho(u,q_I)\le g_I\) or \(p_Ig_I - f_I \le p_I\rho(u,q_0)\). Under this case, we set
\begin{equation}\label{eq:proposition2_1}
\displaystyle \tilde{p}_I = \frac{f_I + \big[\rho(u,q_I)-g_I\big]^+p_I}{\rho(u,q_I)}.
\end{equation}
Next, we show that the ISP has the same revenue and data load under the two pricing strategies \(\theta_I\) and \(\tilde{\theta}_I\). 

If \(q_I\ge q_0\), all users would choose the free provider no matter the ISP adopts the pricing strategy \(\theta_I\) or \(\tilde{\theta}_I\). Thus we have that \(R_I(\tilde{\theta}_I,q_I,u) = R_I(\theta_I,q_I,u) = 0\) and \(d_I(\tilde{\theta}_I,q_I,u) = d_I(\theta_I,q_I,u)= 0\). 


If \(q_I< q_0\) and \(\rho(u,q_I)\le g_I\), by Equation (\ref{eq:actual demand}), it satisfies that
\begin{equation}\label{eq:proposition2_2}
y^*(\phi,\tilde{\theta}_I,q_I) = y^*(\phi,\theta_I,q_I) = \rho(u,q_I) \le g_I
\end{equation}
for any user type \(\phi = (u,v), \forall v\in [0,1]\). Therefore, by Equations (\ref{eq:utility function}) and (\ref{eq:proposition2_1}), we have that
\begin{equation*}
\pi\big(y^*(\phi,\tilde{\theta}_I,q_I),v,\tilde{\theta}_I\big) = \pi\big(y^*(\phi,\theta_I,q_I),v,\theta_I\big) = v\rho(u,q_I) - f_I.
\end{equation*}
By Definition \ref{def:prefer}, we know that \(V_I(\tilde{\theta}_I,q_I,u) = V_I(\theta_I,q_I,u)\) and thus we have
\[R_I(\tilde{\theta}_I,q_I,u) = F'_u(u)\int_{V_I(\tilde{\theta}_I,q_I,u)} f_I dF_v= F'_u(u)\int_{V_I(\theta_I,q_I,u)} f_I d\mu = R_I(\theta_I,q_I,u)\]
by Equation (\ref{eq:demand-based revenue}), and 
\[d_I(\tilde{\theta}_I,q_I,u) = F'_u(u)\int_{V_I(\tilde{\theta}_I,q_I,u)} \rho(u,q_I) dF_v= F'_u(u)\int_{V_I(\theta_I,q_I,u)} \rho(u,q_I) dF_v = d_I(\theta_I,q_I,u)\]
by Equations (\ref{eq:demand-based load}) and (\ref{eq:proposition2_2}).

If \(q_I< q_0\) and \(\rho(u,q_I)> g_I\) and \(p_Ig_I - f_I \le p_I\rho(u,q_0)\), for any user type \(\phi = (u,v), \forall v< p_I\), it satisfies that \(y^*(\phi,\theta_I,q_I) = g_I\) and \(y^*(\phi,\tilde{\theta}_I,q_I) = \rho(u,q_I)\) by Equation (\ref{eq:actual demand}). Thus we have that 
\begin{align*}
&\pi\big(y^*(\phi,\theta_I,q_I),v,\theta_I\big) = vg_I - f_I \le p_Ig_I - f_I \le v\rho(u,q_0) \quad \text{and} \\
&\pi\big(y^*(\phi,\tilde{\theta}_I,q_I),v,\tilde{\theta}_I\big) = v\rho(u,q_I) - f_I - \big[\rho(u,q_I)-g_I\big]p_I\le v\rho(u,q_I) -p_I\rho(u,q_I)+p_I\rho(u,q_0)\le v\rho(u,q_0)
\end{align*}
by Equation (\ref{eq:utility function}). It implies that the user type \(\phi = (u,v), \forall v< p_I\) would not choose the ISP no matter it adopts the pricing strategy \(\theta_I\) or \(\tilde{\theta}_I\). Thus, we know that \(V_I(\tilde{\theta}_I,q_I,u), V_I(\theta_I,q_I,u) \subseteq [p_I,V]\). Furthermore, for any user type \(\phi = (u,v), \forall v\ge p_I\), it satisfies that 
\[y^*(\phi,\tilde{\theta}_I,q_I) = y^*(\phi,\theta_I,q_I) = \rho(u,q_I)\ge g_I\]
by Equation (\ref{eq:actual demand}).
Therefore, by Equations (\ref{eq:utility function}) and (\ref{eq:proposition2_1}), for any user type \(\phi = (u,v), \forall v\ge p_I\), we have that
\[\pi\big(y^*(\phi,\tilde{\theta}_I,q_I),v,\tilde{\theta}_I\big) = \pi\big(y^*(\phi,\theta_I,q_I),v,\theta_I\big) = v\rho(u,q_I) - f_I - \big[\rho(u,q_I)-g_I\big]p_I\]
implying that \(V_I(\tilde{\theta}_I,q_I,u) = V_I(\theta_I,q_I,u)\). Then by Equations (\ref{eq:demand-based revenue}) and (\ref{eq:demand-based load}), we know that
\begin{align*}
&R_I(\tilde{\theta}_I,q_I,u) = F'_u(u)\int_{V_I(\tilde{\theta}_I,q_I,u)} f_I + \big[\rho(u,q_I)-g_I\big]p_I dF_v= F'_u(u)\int_{V_I(\theta_I,q_I,u)} f_I + \big[\rho(u,q_I)-g_I\big]p_I dF_v = R_I(\theta_I,q_I,u) \\
&d_I(\tilde{\theta}_I,q_I,u) = F'_u(u)\int_{V_I(\tilde{\theta}_I,q_I,u)} \rho(u,q_I) dF_v= F'_u(u)\int_{V_I(\theta_I,q_I,u)} \rho(u,q_I) dF_v = d_I(\theta_I,q_I,u).
\end{align*}

The second case is when \(q_I< q_0\) and \(\rho(u,q_I)> g_I\) and \(p_Ig_I-f_I > p_I\rho(u,q_0)\). Under this case, by Definition \ref{def:prefer} and Equation (\ref{eq:demand-based revenue}), when the ISP adopts the pricing strategy \(\theta_I\), the set of the values of the users who choose the ISP and the revenue of the ISP are
\begin{align}\label{eq:proposition2_3}
V_I(\theta_I,q_I,u) = \left\{v\in [0,1]: v\ge \frac{f_I}{g_I - \rho(u,q_0)}\right\} \ \ \text{and}\ \  R_I(\theta_I,q_I,u) = F'_u(u)\int_{V_I(\theta_I,q_I,u)} t\big(y^*((u,v),\theta_I,q_I),\theta_I\big) dF_v
\end{align}
respectively.  When the ISP adopts the pricing strategy \(\tilde{\theta}_I\), the set of the values of the users who choose the ISP and the revenue of the ISP are
\begin{align}\label{eq:proposition2_4}
V_I(\tilde{\theta}_I,q_I,u) = \left\{v\in [0,1]: v\ge \frac{\rho(u,q_I)\tilde{p}_I}{\rho(u,q_I) - \rho(u,q_0)}\right\} \quad \text{and} \quad R_I(\tilde{\theta}_I,q_I,u) = F'_u(u)\int_{V_I(\tilde{\theta}_I,q_I,u)} \tilde{p_I}\rho(u,q_I) dF_v.
\end{align}

We set \(\tilde{p}_I\) as the solution of the equlations:
\begin{equation}\label{equation:equation set}
\begin{cases}
\displaystyle\left(\frac{1}{\beta+1}\right)^{\frac{1}{\beta}}\frac{1}{M} \le \tilde{p}_I \le \frac{1}{M}\\
\displaystyle R_I (\tilde{\theta}_I,q_I,u) = R_I(\theta_I,q_I,u)
\end{cases}
\quad \text{where} \quad M\triangleq \frac{\rho(u,q_I)}{\rho(u,q_I) - \rho(u,q_0)}.
\end{equation}

Next, we first show that Equations (\ref{equation:equation set}) has a unique solution. Based on Assumption \ref{ass:model parameters}, the distribution function of user valuation is \(F_v(v) = v^{\beta}\). We define a continuous function \(y(x) \triangleq \int_{x}^{1}xdF_v(v) = x(1-x^{\beta}), \forall x\in [0,1]\). It is increasing and concave in \(x\in [0,(\frac{1}{\beta+1})^{\frac{1}{\beta}}]\), decreasing in \(x\in [(\frac{1}{\beta+1})^{\frac{1}{\beta}}, 1]\) and satisfies that
\(0 = y(1) \le y(x) \le y\big((\frac{1}{\beta+1})^{\frac{1}{\beta}}\big)\) for \(\forall x\in [0,1].\)
By Equation (\ref{eq:proposition2_4}), we have that
\begin{align}\label{eq:proposition2_6}
R_I(\tilde{\theta}_I,q_I,u) = \int_{M\tilde{p}_I}^{1} \rho(u,q_I)\tilde{p}_I dF_vF'_u(u) = \big[\rho(u,q_I) - \rho(u,q_0)\big] y(M\tilde{p}_I) F'_u(u)
\end{align}
implying that \(R_I(\tilde{\theta}_I,q_I,u)\) is decreasing in \(\tilde{p}_I\in \left[\left(\frac{1}{\beta+1}\right)^{\frac{1}{\beta}}\frac{1}{M}, \frac{1}{M}\right]\).
By Equation (\ref{eq:proposition2_3}), we have that
\begin{align}\label{eq:proposition2_7}
R_I(\theta_I,q_I,u) &= \left\{\int_{\frac{f_I}{g_I-\rho(u,q_0)}}^{p_I} f_IdF_v + \int_{p_I}^1 f_I + \big[\rho(u,q_I)-g_I\big]p_I dF_v\right\}F'_u(u)\notag\\
&=\left\{\big[g_I-\rho(u,q_0)\big]\int_{\frac{f_I}{g_I-\rho(u,q_0)}}^{1}\frac{f_I}{g_I-\rho(u,q_0)}dF_v + \big[\rho(u,q_I) - g_I\big]\int_{p_I}^1 p_I dF_v\right\}F'_u(u)\notag\\
&=\left\{\big[g_I-\rho(u,q_0)\big]y\left(\frac{f_I}{g_I-\rho(u,q_0)}\right) + \big[\rho(u,q_I) - g_I\big]y(p_I)\right\}F'_u(u) \\
&\le \big[\rho(u,q_I) - \rho(u,q_0)\big]y\left((\frac{1}{\beta+1})^{\frac{1}{\beta}}\right)F'_u(u)\notag
\end{align}
from which it satisfies
\[R_I\left(\big(0,0,(\frac{1}{\beta+1})^{\frac{1}{\beta}}\!\frac{1}{M}\big),q_I,u\right) = \big[\rho(u,q_I) - \rho(u,q_0)\big]y\left((\frac{1}{\beta+1})^{\frac{1}{\beta}}\right)F'_u(u) \ge R_I(\theta_I,q_I,u) \ge 0 =
R_I\left(\big(0,0,\frac{1}{M}\big),q_I,u\right) \]
implying that Equations (\ref{equation:equation set}) must have a unique solution.

We then show that \(d_I(\theta_I,q_I,u) \ge d_I(\tilde{\theta}_I,q_I,u)\) under two situations: (i) \(p_I\le M\tilde{p}_I\) and (ii) \(\ p_I> M \tilde{p}_I\).

For the situation (i), it satisfies that \(f_I/\big[g_I-\rho(u,q_0)\big] < p_I\le M\tilde{p}_I\). By Equation (\ref{eq:demand-based load}), we can deduce that 
\begin{equation*}
d_I(\tilde{\theta}_I,q_I,u) = \int_{M\tilde{p}_I}^{1} \rho(u,q_I) dF_vF'_u(u) \le \int_{p_I}^1 \rho(u,q_I) dF_vF'_u(u) +\int_{\frac{f_I}{g_I-\rho(u,q_0)}}^{p_I} g_IdF_vF'_u(u) = d_I(\theta_I,q_I,u).
\end{equation*}

For the situation (ii), we first define continuous functions \(z(x) = \int_{x}^1dF_v(v) = 1 -x^{\beta}\) and \(w(x) = x(1-x)^{\frac{1}{\beta}}, \forall x\in[0,1]\) and it satisfies that
\begin{equation}\label{eq:proposition2_5}
w\big(z(x)\big) = z(x)\big(1-z(x)\big)^{\frac{1}{\beta}} = x(1-x^\beta) = y(x).
\end{equation}
By Equations (\ref{equation:equation set}), (\ref{eq:proposition2_6}), (\ref{eq:proposition2_7}) and (\ref{eq:proposition2_5}), we have that 
\begin{equation}\label{eq:dragon}
\big[\rho(u,q_I)-\rho(u,q_0)\big]w\big(z(M\tilde{p}_I)\big) = \big[\rho(u,q_I) - g_I\big]w\big(z(p_I)\big)+\big[g_I-\rho(u,q_0)\big]w\left(z\Big(\frac{f_I}{g_I-\rho(u,q_0)}\Big)\right).
\end{equation}

If \(\displaystyle\frac{f_I}{g_I-\rho(u,q_0)} \ge (\frac{1}{\beta+1})^{\frac{1}{\beta}}\), we have \(\displaystyle 0\le z(p_I) \le z(M\tilde{p}_I) \le z\Big(\frac{f_I}{g_I-\rho(u,q_0)}\Big) \le z\Big((\frac{1}{\beta+1})^{\frac{1}{\beta}}\Big) = \frac{\beta}{\beta+1}\). 
Because \(w(z)\) is increasing and concave in \(z\in [0,\frac{\beta}{\beta+1}]\), by Equation (\ref{eq:dragon}) and Jensen's inequality, we can deduce that 
\begin{align*}
\big[\rho(u,q_I)-\rho(u,q_0)\big]z\big(M\tilde{p}_I\big) \le \big[\rho(u,q_I) - g_I\big]z(p_I)+\big[g_I-\rho(u,q_0)\big]z\left(\frac{f_I}{g_I-\rho(u,q_0)}\right).
\end{align*}

If \(\displaystyle\frac{f_I}{g_I-\rho(u,q_0)} < (\frac{1}{\beta+1})^{\frac{1}{\beta}}\), there must exist a value \(k_I\in \big[(\frac{1}{\beta+1})^{\frac{1}{\beta}},1\big]\) which satisfies that \(w\big(z(k_I)\big) = w\Big(z\big(\frac{f_I}{g_I-\rho(u,q_0)}\big)\Big),\) because \(w\big(z(x)\big)\) is continiously decreasing in \(x\in \big[(\frac{1}{\beta+1})^{\frac{1}{\beta}}, 1\big]\) and satisfies that \(w\Big(z\big((\frac{1}{\beta+1})^{\frac{1}{\beta}}\big)\Big)\ge w\big(z(\frac{f_I}{g_I-\rho(u,q_0)})\big) \ge w\big(z(1)\big).\)
Then we have  \(0\le z(p_I)\le z(M\tilde{p}_I)\le z(k_I) \le \frac{\beta}{\beta+1}\) and \(z(k_I) \le z(\frac{f_I}{g_I-\rho(u,q_0)})\). Because \(w(z)\) is increasing and concave in \(z\in [0,\frac{\beta}{\beta+1}]\), by Equation (\ref{eq:dragon}) and Jensen's inequality, we can deduce that 
\begin{align*}
\big[\rho(u,q_I)-\rho(u,q_0)\big]z\big(M\tilde{p}_I\big) &\le \big[\rho(u,q_I) - g_I\big]z(p_I)+\big[g_I-\rho(u,q_0)\big]z(k_I) \\
&\le \big[\rho(u,q_I) - g_I\big]z(p_I)+\big[g_I-\rho(u,q_0)\big]z\left(\frac{f_I}{g_I-\rho(u,q_0)}\right).
\end{align*}
Therefore, combining with Equation (\ref{eq:demand-based load}), we have that
\begin{align*}
& d_I(\tilde{\theta}_I,q_I,u)  = \!\! \int_{M\tilde{p}_I}^{1} \rho(u,q_I) dF_vF'_u(u) = \rho(u,q_I)z\big(M\tilde{p}_I\big)F'_u(u) = \Big\{\!\big[\rho(u,q_I)-\rho(u,q_0)\big]z\big(M\tilde{p}_I\big) + \rho(u,q_0)z\big(M\tilde{p}_I\big)\!\Big\}F'_u(u)\\
& \le \left\{\big[\rho(u,q_I) - g_I\big]z(p_I)+\big[g_I-\rho(u,q_0)\big]z\left(\frac{f_I}{g_I-\rho(u,q_0)}\right) + \rho(u,q_0)z\left(\frac{f_I}{g_I-\rho(u,q_0)}\right)\right\}F'_u(u)\\
&= \left\{\int_{\frac{f_I}{g_I-\rho(u,q_0)}}^{p_I} g_IdF_v + \int_{p_I}^1 \rho(u,q_I)dF_v\right\}F'_u(u)= d_I(\theta_I,q_I,u).\tag*{\QEDA}
\end{align*}

\textbf{Proof of Lemma \ref{lemma:equilibrium expansion}:} This lemma can be proofed by the similar way with Theorem \ref{theorem:equilibrium}. In particular, for any demand-based pricing \(\theta_I(u)\), we can show the existence of the equilibrium based on the Brouwer fixed-point theorem and the uniqueness of the equilibrium by contradiction.
\QEDA\\

\textbf{Proof of Proposition \ref{proposition:grained comparison extension}:} By Proposition \ref{proposition:grained comparison}, for any \(u_0\in [0,1]\) and the fixed two-part pricing \(\theta_I^{u_0} \triangleq \theta_I(u_0)\), there always exists a fixed pay-as-you-go pricing \(\tilde{\theta}_I^{u_0}\) satisfying that
\begin{align}\label{equation:better efficiency}
\begin{cases}
R_I\Big(\tilde{\theta}_I^{u_0},q_I\big(\theta_I(u)\big),u_0\Big) = R_I\Big(\theta_I^{u_0},q_I\big(\theta_I(u)\big),u_0\Big)\vspace{0.05in}\\
d_I\Big(\tilde{\theta}_I^{u_0},q_I\big(\theta_I(u)\big),u_0\Big) \le d_I\Big(\theta_I^{u_0},q_I\big(\theta_I(u)\big),u_0\Big).
\end{cases}
\end{align}
We conduct a demand-based pay-as-you-go pricing \(\tilde{\theta}_I(u)\) which satisfies \(\tilde{\theta}_I(u_0) = \tilde{\theta}_I^{u_0}\) for any \(u_0\in [0,1]\). Thus we have that
\begin{equation}\label{eq:proposition3_1}
R_I\Big(\tilde{\theta}_I(u),q_I\big(\theta_I(u)\big)\Big) = \int_0^1R_I\Big(\tilde{\theta}_I^{u_0},q_I\big(\theta_I(u)\big),u_0\Big)du_0=\int_0^1R_I\Big(\theta_I^{u_0},q_I\big(\theta_I(u)\big),u_0\Big)du = R_I\Big(\theta_I(u),q_I\big(\theta_I(u)\big)\Big),
\end{equation}
\begin{equation}\label{equation:aggregate demand}
d_I\Big(\tilde{\theta}_I(u),q_I\big(\theta_I(u)\big)\Big) = \int_0^1d_I\Big(\tilde{\theta}_I^{u_0},q_I\big(\theta_I(u)\big),u_0\Big)du_0 \le \int_0^1d_I\Big(\theta_I^{u_0},q_I\big(\theta_I(u)\big),u_0\Big)du = d_I\Big(\theta_I(u),q_I\big(\theta_I(u)\big)\Big).
\end{equation}

Next, we show that \(q_I\big(\tilde{\theta}_I(u)\big)\le q_I\big(\theta_I(u)\big)\) by contradiction. Suppose that \(q_I\big(\tilde{\theta}_I(u)\big)> q_I\big(\theta_I(u)\big)\). By Equations (\ref{eq:max utility}) and (\ref{eq:actual demand}), any user's optimal utility and usage are both non-increasing in the congestion level, and thus the ISP's data load under the congestion \(q_I\big(\theta_I(u)\big)\) is no smaller than that under the congestion \(q_I\big(\tilde{\theta}_I(u)\big)\), i.e., \(d_I\big(\tilde{\theta}_I(u),q_I\big(\tilde{\theta}_I(u)\big)\big) \le d_I\big(\tilde{\theta}_I(u),q_I\big(\theta_I(u)\big)\big).\) Furthermore, Combining with Inequality (\ref{equation:aggregate demand}), we have that 
\[d_I\Big(\tilde{\theta}_I(u),q_I\big(\tilde{\theta}_I(u)\big)\Big) \le d_I\Big(\tilde{\theta}_I(u),q_I\big(\theta_I(u)\big)\Big) \le d_I\Big(\theta_I(u),q_I\big(\theta_I(u)\big)\Big).\]
By Assumption \ref{ass:congestion_function}, we can deduce that
\[q_I\big(\tilde{\theta}_I(u)\big) = Q_I\bigg(d_I\Big(\tilde{\theta}_I(u),q_I\big(\tilde{\theta}_I(u)\big)\Big),c_I\bigg) \le Q_I\bigg(d_I\Big(\theta_I(u),q_I\big(\theta_I(u)\big)\Big),c_I\bigg) = q_I\big(\theta_I(u)\big)\]
which is contradictory with the supposition. Thus it satisfies that \(q_I\big(\tilde{\theta}_I(u)\big)\le q_I\big(\theta_I(u)\big)\).

Finally, because any user's optimal utility and usage are both non-increasing in the congestion level, the ISP's revenue is also non-increasing in the congestion level, combining with Inequality (\ref{eq:proposition3_1}), we have that
\[R_I\Big(\tilde{\theta}_I(u),q_I\big(\tilde{\theta}_I(u)\big)\Big) \ge R_I\Big(\tilde{\theta}_I(u),q_I\big(\theta_I(u)\big)\Big) = R_I\Big(\theta_I(u),q_I\big(\theta_I(u)\big)\Big).\tag*{\QEDA}\]

\textbf{Proof of Proposition \ref{proposition:grained explanation}:} It is obvious that if \(\tilde{\theta}_I(u) \triangleq \big(0,0,\tilde{p}_I(u)\big)\) where \(\tilde{p}_I(u) = \tilde{p}_I\) for all \(u\in [0,1]\) is a revenue-optimal demand-based pay-as-you-go pricing, \(\tilde{\theta}_I \triangleq \big(0,0,\tilde{p}_I\big)\) must be a revenue-optimal fixed pay-as-you-go pricing. Thus we focus on proving that if \(\tilde{\theta}_I \triangleq \big(0,0,\tilde{p}_I\big)\) is a revenue-optimal fixed pay-as-you-go pricing, \(\tilde{\theta}_I(u) \triangleq \big(0,0,\tilde{p}_I(u)\big)\) where \(\tilde{p}_I(u) = \tilde{p}_I\) for all \(u\in [0,1]\) must be a revenue-optimal demand-based pay-as-you-go pricing.
To prove this, we only need to show that any revenue-optimal demand-based pay-as-you-go pricing \(\tilde{\theta}_I(u)\) must satisfy that \(\tilde{p}_I(u)\) is a constant function of \(u\in [0,1]\).
Under Assumption \ref{ass:model parameters}, we have \(F_v(x) = x^{\beta}\). We denote that \(h(u) \triangleq \big(\tilde{p}_I(u)\big)^{\beta}, \forall u\in [0,1]\). Thus, the equilibrium congestion and revenue of the ISP under the pricing strategy \(\tilde{\theta}_I(u)\) are functions of \(h(u)\). We denote
\[s\big(h(u)\big) \triangleq q_I\big(\tilde{\theta}_I(u)\big)\quad \text{and} \quad T\big(h(u)\big) \triangleq R_I\Big(\tilde{\theta}_I(u),q_I\big(\tilde{\theta}_I(u)\big)\Big).\]
Because \(\tilde{\theta}_I(u)\) is a revenue-optimal pricing strategy, \(h(u)\) is an optimal solution of the problem:
\begin{align*}
&\text{Maximize} \quad T\big(x(u)\big)\\
&\text{subject to} \quad \,x(u)\ge 0 \quad \text{for} \ \ \forall u\in [0,1].\notag
\end{align*}
We define a function \(H(\epsilon) \triangleq T\big(h(u) + \epsilon \eta(u)\big)\) where \(\eta(u)\) is a disturbing function. Because \(T\big(h(u)\big) \ge T\big(h(u) + \epsilon \eta(u)\big)\), we have the first order condition
\begin{align}\label{equation:revenue derive}
\frac{dH(\epsilon)}{d\epsilon}\big|_{\epsilon=0} = 0
\end{align}

By Equation (\ref{eq:demand-based load}) and (\ref{eq:demand-based load2}), the ISP's data load under the pricing strategy \(\tilde{\theta}_I(u)\) is
\begin{equation*}
\begin{aligned}
& d_I\Big(\tilde{\theta}_I(u),q_I\big(\tilde{\theta}_I(u)\big)\Big)  = \int_0^1\int_{K\big(q_I(\tilde{\theta}_I(u))\big)\tilde{p}_I(u_0)}^1\rho\Big(u_0,q_I\big(\tilde{\theta}_I(u)\big)\Big)dF_v(v)dF_u(u_0)\\
& = \int_0^1\int^1_{K\big(s(h(u))\big)\big(h(u_0)\big)^{\frac{1}{\beta}}}u_0 e^{-s(h(u))} dF_v(v)dF_u(u_0) = \int_0^1\left\{1-\Big[K\big(s(h(u))\big)\Big]^\beta h(u_0)\right\} u_0dF_u(u_0)\cdot e^{-s(h(u))}.
\end{aligned}
\end{equation*}
where \(K(x) \triangleq e^{-x}/(e^{-x}-e^{-q_0})\). By Assumption \ref{ass:model parameters}, we have \(d_I = c_I q_I\) and thus it satisfies that
\begin{equation}\label{eq:proposition4_1}
\int_0^1\left\{1-\Big[K\big(s(h(u))\big)\Big]^\beta h(u_0)\right\} u_0dF_u(u_0)\cdot e^{-s(h(u))} = c_I s\big(h(u)\big).
\end{equation}
We define a function \(G(x) = [\int_0^1 u_0dF_u(u_0) - c_Ixe^x]/[K(x)]^\beta,\forall x\in [0,q_0)\) and \(G(x)\) is decreasing in \(x\). By Equation (\ref{eq:proposition4_1}), it satisfies that
\[G\Big(s\big(h(u)\big)\Big) = \frac{\int_0^1 u_0dF_u(u_0) - c_Is\big(h(u)\big)e^{s\big(h(u)\big)}}{\Big[K\big(s(h(u))\big)\Big]^{\beta}} = \int_0^1h(u_0) u_0dF_u(u_0)\]
from which we have that 
\[s\big(h(u)\big) = G^{-1}\left(\int_0^1h(u_0)u_0dF_u(u_0)\right).\]
Furthermore, for any disturbing function \(\eta(u)\) satisfying \(\int_0^1 \eta(u_0)u_0dF_u(u_0) = 0\), we have that
\begin{align}\label{equation:congestion derive}
\frac{ds\big(h(u)+\epsilon\eta(u)\big)}{d\epsilon}\Big|_{\epsilon=0} = (G^{-1})' \left(\int_0^1h(u_0)u_0dF_u(u_0)\right)\cdot \int_0^1 \eta(u_0)u_0dF_u(u_0) = 0
\end{align}
By Equations (\ref{eq:demand-based revenue}) and (\ref{eq:demand-based load2}), the ISP's revenue under the pricing strategy \(\tilde{\theta}_I(u)\) is
\begin{equation}\label{eq:proposition4_2}
\begin{aligned}
T\big(h(u)\big) = R_I\Big(\tilde{\theta}_I(u),q_I\big(\tilde{\theta}_I(u)\big)\Big)  & = \int_0^1\int_{K\big(q_I(\tilde{\theta}_I(u))\big)\tilde{p}_I(u_0)}^1\tilde{p}_I(u_0)\rho\Big(u_0,q_I\big(\tilde{\theta}_I(u)\big)\Big)dF_v(v)dF_u(u_0)\\
& = \int_0^1\int^1_{K\big(s(h(u))\big)\big(h(u_0)\big)^{\frac{1}{\beta}}}\big[h(u_0)\big]^{\frac{1}{\beta}}u_0 e^{-s(h(u))} dF_v(v)dF_u(u_0)\\
& = \int_0^1\left\{1-\Big[K\big(s(h(u))\big)\Big]^\beta h(u_0)\right\} \big[h(u_0)\big]^{\frac{1}{\beta}}u_0dF_u(u_0)\cdot e^{-s(h(u))}.
\end{aligned}
\end{equation}
We define a function \(W(x) \triangleq x^{\frac{1}{\beta}-1}\left\{1-(\beta+1)x\Big[K\big(s(h(u))\big)\Big]^\beta\right\}.\)
By Equations (\ref{equation:revenue derive}), (\ref{equation:congestion derive}), and (\ref{eq:proposition4_2}), we can deduce that
\begin{equation*}
\begin{aligned}
&\frac{dH(\epsilon)}{d\epsilon}\Big|_{\epsilon=0} = \frac{dT\big(h(u) + \epsilon \eta(u)\big)}{d\epsilon}\Big|_{\epsilon=0} = \int_0^1\left\{1-\Big[K\big(s(h(u))\big)\Big]^\beta h(u_0)\right\} \frac{1}{\beta}\big[h(u_0)\big]^{\frac{1}{\beta}-1}\eta(u_0)u_0dF_u(u_0)\cdot e^{-s(h(u))} \\
& - \int_0^1\Big[K\big(s(h(u))\big)\Big]^\beta \eta(u_0)\big[h(u_0)\big]^{\frac{1}{\beta}}u_0dF_u(u_0)\cdot e^{-s(h(u))} = \frac{1}{\beta}\int_0^1 \eta(u_0)u_0W\big(h(u_0)\big)dF_u(u_0)\cdot e^{-s\big(h(u)\big)} = 0.
\end{aligned}
\end{equation*}
Thus, we have \(\int_0^1 \eta(u_0)u_0W\big(h(u_0)\big)dF_u(u_0) = 0\) if \(\int_0^1 \eta(u_0)u_0dF_u(u_0) = 0\).

We make a substitution \(\xi(u_0) = u_0F'_u(u_0)\eta(u_0)\). Then for any function \(\xi(u)\) satisfying \(\int_0^1\xi(u_0)du_0 = 0\), we have \(\int_0^1\xi(u_0)W\big(h(u_0)\big)du_0 = 0\). Based on this condition, we next prove that \(W\big(h(u)\big)\) is a constant function of \(u\in [0,1]\) by contradiction. Suppose \(W\big(h(u)\big)\) is not a constant function. Then there must exist \(\epsilon>0\) and \(u_1,u_2\in (0,1)\) satisfying that \(\big|W\big(h(u_1)\big) - W\big(h(u_2)\big)\big|>\epsilon\). Because we assumed that \(\tilde{p}_I(u)\) is a continuous function on \([0,1]\), \(W\big(h(u)\big)\) is also continuous on \([0,1]\). Furthermore, it must be uniformly continuous on \([0,1]\). By the uniform continuity, there must exist \(\delta\) satisfying \(\big|W\big(h(\hat{u})\big) - W\big(h(u)\big)\big|<\frac{\epsilon}{3}\), \(\forall |\hat{u}-u|<\delta\). We conduct the function \(\xi(u)\) as follows:
\begin{equation*}
\xi(u)=
\begin{cases}
1 \ \ \ &\text{if} \ \ |u-u_1|<\delta,\\
-1 \ \ \ &\text{if} \ \ |u-u_2|<\delta,\\
0 \ \ \ &\text{otherwise}.
\end{cases}
\end{equation*}
It is obvious that \(\int_0^1 \xi(u_0)du_0 = 0\). However,
\begin{equation*}
\begin{aligned}
\Big|\int_0^1 \xi(u_0)W\big(h(u_0)\big)du_0\Big| &= \Big|\int_{u_1-\delta}^{u_1+\delta} W\big(h(u_0)\big)du_0 - \int_{u_2-\delta}^{u_2+\delta} W\big(h(u_0)\big)du_0\Big|\\
&\ge \Big(\Big|(W\big(h(u_1)\big)-W\big(h(u_2)\big)\Big| - \frac{2\epsilon}{3}\Big)\cdot 2\delta \ge \frac{2\epsilon\delta}{3}>0.
\end{aligned}
\end{equation*}
This is contradictory with \(\int_0^1 \xi(u_0)W\big(h(u_0)\big)du_0 = 0\) under the condition \(\int_0^1\xi(u_0)du_0 = 0\). Thus, \(W\big(h(u)\big)\) is a constant function of \(u\).
Furthermore, \(h(u)\) is a constant function of \(u\) and thus \(\tilde{p}_I(u)\) is also a constant function of \(u\). Therefore, Proposition \ref{proposition:grained comparison extension} is proved.
\QEDA\\

{\setlength{\bibsep}{0.0pt}
\bibliographystyle{ieeetr}
\bibliography{tex}

\begin{thebibliography}{10}

\bibitem{xin2005role}
X.~Wang, R.~T.~B. Ma, and Y.~Xu, ``The role of data cap in optimal two-part
  network pricing,'' {\em Proceedings of the 24th International World Wide Web
  Conference (WWW)}, 2015.

\bibitem{wang2015sdp}
X.~Wang, R.~T.~B. Ma, and Y.~Xu, ``The role of data cap in two-part pricing
  under market competition,'' {\em Proceedings of the 4th Smart Data Pricing
  Workshop (SDP)}, 2015.

\bibitem{wang2016network}
X.~Wang, R.~T.~B. Ma, and Y.~Xu, ``The role of data cap in two-part pricing
  under market competition,'' {\em IEEE Network}, vol.~30(2), 2016.

\bibitem{odlyzko2001internet}
A.~Odlyzko, ``Internet pricing and the history of communications,'' {\em
  Computer networks}, vol.~36, no.~5, pp.~493--517, 2001.

\bibitem{labovitz2011internet}
C.~Labovitz, S.~Iekel-Johnson, D.~McPherson, J.~Oberheide, and F.~Jahanian,
  ``Internet inter-domain traffic,'' {\em ACM SIGCOMM Computer Communication
  Review}, vol.~41, no.~4, pp.~75--86, 2011.

\bibitem{nabipay2011flat}
P.~Nabipay, A.~Odlyzko, and Z.-L. Zhang, ``Flat versus metered rates, bundling,
  and ``bandwidth hogs'','' in {\em 6th Workshop on the Economics of Networks,
  Systems, and Computation}, 2011.

\bibitem{morgan2011pricing}
M.~Morgan, ``{Pricing schemes key in LTE future},'' {\em Telecomasia. net,
  September}, vol.~12, 2011.

\bibitem{segall2011verizon}
L.~Segall, ``Verizon ends unlimited data plan,'' {\em CNN Money. July}, vol.~6,
  2011.

\bibitem{taylor2011t}
P.~Taylor, ``{AT\&T imposes usage caps on fixed-line broadband},'' {\em
  Financial Times. March}, vol.~14, 2011.

\bibitem{ATTunlimited}
``{AT\&T's unlimited data plan}.''
  \url{https://www.att.com/plans/unlimited-data-plans.html}.

\bibitem{Verizonunlimited}
``{Verizon's unlimited data plan}.''
  \url{https://www.verizonwireless.com/plans/verizon-plan}.

\bibitem{edell1999providing}
R.~Edell and P.~Varaiya, ``{Providing Internet access: What we learn from
  INDEX},'' {\em IEEE Network}, vol.~13, no.~5, pp.~18--25, 1999.

\bibitem{dai2013design}
W.~Dai and S.~Jordan, ``Design and impact of data caps,'' in {\em IEEE Global
  Communications Conference}, pp.~1650--1656, 2013.

\bibitem{open2013policy}
O.~I.~A. Committee {\em et~al.}, ``Policy issues in data caps and usage-based
  pricing,'' {\em Economic Impacts of Open Internet Frameworks Working Group},
  p.~13, 2013.

\bibitem{schatz10fcc}
A.~Schatz and S.~E. Ante, ``{FCC chief backs usage-based broadband pricing},''
  {\em Wall Street Journal}, December 2, 2010.

\bibitem{oi1971disneyland}
W.~Y. Oi, ``A disneyland dilemma: Two-part tariffs for a mickey mouse
  monopoly,'' {\em The Quarterly Journal of Economics}, pp.~77--96, 1971.

\bibitem{calem1984multiproduct}
P.~S. Calem and D.~F. Spulber, ``Multiproduct two part tariffs,'' {\em
  International Journal of Industrial Organization}, vol.~2, no.~2,
  pp.~105--115, 1984.

\bibitem{littlechild1975two}
S.~C. Littlechild, ``Two-part tariffs and consumption externalities,'' {\em The
  Bell Journal of Economics}, pp.~661--670, 1975.

\bibitem{scotchmer1985two}
S.~Scotchmer, ``Two-tier pricing of shared facilities in a free-entry
  equilibrium,'' {\em The Rand Journal of Economics}, pp.~456--472, 1985.

\bibitem{dai2013isp}
W.~Dai and S.~Jordan, ``How do {ISP} data caps affect subscribers?,'' {\em
  Telecommunications Policy Research Conference TPRC}, 2013.

\bibitem{odlyzko2012know}
A.~Odlyzko, B.~S. Arnaud, E.~Stallman, and M.~Weinberg, ``{Know your limits:
  Considering the role of data caps and usage based billing in Internet access
  service},'' {\em Public Knowledge}, April 23, 2012.

\bibitem{chetty2012you}
M.~Chetty, R.~Banks, A.~Brush, J.~Donner, and R.~Grinter, ``You're capped:
  understanding the effects of bandwidth caps on broadband use in the home,''
  in {\em Proceedings of the ACM SIGCHI Conference on Human Factors in
  Computing Systems}, pp.~3021--3030, 2012.

\bibitem{Poularakis2014pricing}
K.~Poularakis, I.~Pefkianakis, J.~Chandrashekar, and L.~Tassiulas, ``Pricing
  the last mile: Data capping for residential broadband,'' in {\em Proceedings
  of the 10th ACM International on Conference on emerging Networking
  Experiments and Technologies}, pp.~295--306, 2014.

\bibitem{hande2010pricing}
P.~Hande, M.~Chiang, R.~Calderbank, and J.~Zhang, ``Pricing under constraints
  in access networks: Revenue maximization and congestion management,'' in {\em
  Proceedings of the IEEE INFOCOM}, pp.~1--9, 2010.

\bibitem{li2009revenue}
S.~Li, J.~Huang, and S.-Y.~R. Li, ``Revenue maximization for communication
  networks with usage-based pricing,'' in {\em IEEE Global Telecommunications
  Conference}, pp.~1--6, 2009.

\bibitem{basar2002revenue}
T.~Basar and R.~Srikant, ``Revenue-maximizing pricing and capacity expansion in
  a many-users regime,'' in {\em Proceedings of the IEEE INFOCOM},
  pp.~294--301, 2002.

\bibitem{shen2007optimal}
H.~Shen and T.~Basar, ``Optimal nonlinear pricing for a monopolistic network
  service provider with complete and incomplete information,'' {\em IEEE
  Journal on Selected Areas in Communications}, vol.~25, no.~6, 2007.

\bibitem{chen2016impact}
C.~Chen, R.~A. Berry, M.~L. Honig, and V.~G. Subramanian, ``The impact of
  unlicensed access on small-cell resource allocation,'' in {\em Proceedings of
  the IEEE INFOCOM}, pp.~1--9, 2016.

\bibitem{ma2016usage}
R.~T.~B. Ma, ``Usage-based pricing and competition in congestible network
  service markets,'' {\em IEEE/ACM Transactions on Networking}, vol.~24, no.~5,
  pp.~3084--3097, 2016.

\bibitem{ha2012tube}
S.~Ha, S.~Sen, C.~Joe-Wong, Y.~Im, and M.~Chiang, ``Tube: time-dependent
  pricing for mobile data,'' {\em ACM SIGCOMM Computer Communication Review},
  vol.~42, no.~4, pp.~247--258, 2012.

\bibitem{henderson2001congestion}
T.~Henderson, J.~Crowcroft, and S.~Bhatti, ``{Congestion pricing. Paying your
  way in communication networks},'' {\em IEEE Internet Computing}, vol.~5,
  no.~5, pp.~85--89, 2001.

\bibitem{gizelis2011survey}
C.~A. Gizelis and D.~D. Vergados, ``A survey of pricing schemes in wireless
  networks,'' {\em IEEE Communications Surveys \& Tutorials}, vol.~13, no.~1,
  pp.~126--145, 2011.

\bibitem{chander1989optimal}
P.~Chander and L.~Leruth, ``The optimal product mix for a monopolist in the
  presence of congestion effects: A model and some results,'' {\em
  International Journal of Industrial Organization}, vol.~7, no.~4,
  pp.~437--449, 1989.

\bibitem{reitman1991endogenous}
D.~Reitman, ``Endogenous quality differentiation in congested markets,'' {\em
  The Journal of Industrial Economics}, vol.~39, no.~6, pp.~621--647, 1991.

\bibitem{duan2015pricing}
L.~Duan, J.~Huang, and B.~Shou, ``{Pricing for local and global Wi-Fi
  markets},'' {\em IEEE Transactions on Mobile Computing}, vol.~14, no.~5,
  pp.~1056--1070, 2015.

\bibitem{ma13public}
R.~T.~B. Ma and V.~Misra, ``The public option: a non-regulatory alternative to
  network neutrality,'' {\em IEEE/ACM Transactions on Networking}, vol.~21,
  pp.~1866--1879, Dec 2013.

\bibitem{Gryta15}
T.~Gryta, ``{FCC} votes to allow municipal broadband, overruling two states'
  laws,'' {\em Wall Street Journal}, Feb 26, 2015.

\bibitem{ma2013evolution}
R.~T.~B. Ma, J.~Lui, and V.~Misra, ``{On the evolution of the Internet economic
  ecosystem},'' in {\em Proceedings of the 22nd international conference on
  World Wide Web}, pp.~849--860, 2013.

\bibitem{chau2010viability}
C.-K. Chau, Q.~Wang, and D.-M. Chiu, ``On the viability of paris metro pricing
  for communication and service networks,'' in {\em Proceedings of IEEE
  INFOCOM}, pp.~1--9, 2010.

\bibitem{gibbens2000internet}
R.~Gibbens, R.~Mason, and R.~Steinberg, ``Internet service classes under
  competition,'' {\em IEEE Journal on Selected Areas in Communications},
  vol.~18, no.~12, pp.~2490--2498, 2000.

\bibitem{jain2001analysis}
R.~Jain, T.~Mullen, and R.~Hausman, ``Analysis of paris metro pricing strategy
  for qos with a single service provider,'' in {\em International Workshop on
  Quality of Service}, pp.~44--58, 2001.

\bibitem{chen2015bandwidth}
C.~Chen, R.~A. Berry, M.~L. Honig, and V.~G. Subramanian, ``Bandwidth
  optimization in hetnets with competing service providers,'' in {\em the
  Fourth IEEE Workshop on Smart Data Pricing}, pp.~504--509, 2015.

\bibitem{tremblay2012new}
V.~J. Tremblay and C.~H. Tremblay, {\em New perspectives on industrial
  organization: With contributions from behavioral economics and game theory}.
\newblock Springer Science \& Business Media, 2012.

\end{thebibliography}
}

\end{document}